\documentclass[aps,prl,twocolumn,groupedaddress]{revtex4}

\usepackage[utf8]{inputenc}
\usepackage{amssymb}
\usepackage{amsmath}
\usepackage{amsfonts}
\usepackage{mathrsfs}
\usepackage{color}
\usepackage{dsfont}
\usepackage{graphics, graphicx}
\usepackage{psfrag}
\usepackage{subfigure}
\usepackage{booktabs}
\usepackage{blindtext}
\usepackage{float}
\usepackage{import}
\usepackage{sansmath}

\usepackage[pdftitle={},
            pdfsubject={},
            pdfauthor={Steffen Backes},
            ]{hyperref}

\usepackage{color}
\usepackage{soul}
\setcitestyle{numbers}

\newcommand{\svo}{SrVO$_3$}
\newcommand{\smo}{SrMoO$_3$}

\begin{document}
\author{Steffen Backes$^{1,2,3}$}
\email[]{steffen-backes@g.ecc.u-tokyo.ac.jp}
\author{Hong Jiang$^4$}
\author{Silke Biermann$^{3,5,6,7}$}
\affiliation{$^1$Research Center for Advanced Science and Technology, University of Tokyo, Komaba, Tokyo 153-8904, Japan}
\affiliation{$^2$Center for Emergent Matter Science, RIKEN, Wako, Saitama 351-0198, Japan}
\affiliation{$^3$CPHT, CNRS, École polytechnique, Institut Polytechnique de Paris, 91120 Palaiseau, France}
\affiliation{$^4$College of Chemistry and Molecular Engineering, Peking University, China}
\affiliation{$^5$Coll\`ege de France, 11 place Marcelin Berthelot, 75005 Paris, France}
\affiliation{$^6$European Theoretical Spectroscopy Facility, 91128 Palaiseau, France, Europe}
\affiliation{$^7$Department of Physics, Division of Mathematical Physics, Lund University, Professorsgatan 1, 22363 Lund, Sweden}

\date{\today}


\
\vspace{0.5cm}
\

\title{Diagnostics for plasmon satellites and Hubbard bands in transition metal oxides}

\begin{abstract}
Coulomb correlations between the electrons imprint 
characteristic signatures to the spectral properties
of materials. Among others, they are at the origin of
a rich phenomenology of satellite features, either
stemming from atomic-like multiplets or
from interactions with particle-hole excitations or
plasmons. While in many cases the latter lie at
considerably higher energies than the former,
suggesting clear distinction criteria, this picture
has recently become blurred by indications that 
satellites of different types can coexist in the same 
energy range. It is now generally accepted that
the identification of the nature of spectral features 
is a highly non-trivial task. In this article we propose
a general procedure for tracing the origin of satellites 
of different types within modern \textit{ab initio} 
calculations. As an illustration, we analyze the ternary
transition metal oxides {\svo} and {\smo}, which are
drosophila compounds for the coexistence of Hubbard and 
plasmonic satellites, reconciling previous seemingly
contradictory findings in an unexpected manner.
\end{abstract}

\maketitle

\section{Introduction}

Impressive progress in direct -- and to a much lesser extent
inverse -- photoemission spectroscopy over the last decades
has resulted in a situation where the spectral properties of 
electronic systems have become some of the most commonly probed
experimental properties of materials 
\cite{Damascelli2004,Sobota2021,Zhang2022}.
The main quantity is the spectral function $A(k,\omega)$, which 
encodes information about the possible electron removal and 
addition processes, as probed in direct and inverse photoemission. 
The knowledge of $A(k,\omega)$ in turn is typically synonymous 
with a good first understanding of the behaviour of the material
under a variety of probes, even those not directly encoded in $A$.

In normal metals, the low-energy behaviour is governed
by renormalized quasi-particle bands following the Landau
Fermi liquid paradigm, while in insulators the spectrum
is gapped around the Fermi level. Beyond these elementary
considerations, spectral functions can however display a
whole zoology of different features at intermediate or
high energies (in typical transition metal oxides, in energy
ranges spanning a few tenths to a few tens of eV).

Among the most prominent features in electronic systems with
sizable Coulomb correlations are Hubbard satellites, remnants
of the atomic physics in the material, corresponding to the
atomic multiplets of an isolated atom placed in the crystal
field environment of its surroundings but potentially acquiring
some dispersion due to the periodicity of the crystal.
The energy scales of these multiplet structures are given
by the effective local Coulomb interaction, often parametrized
theoretically in the form of a local Hubbard $U$ (or more 
precisely a Hubbard $U$ matrix including Hund's exchange and
orbital structures) \cite{Hubbard1963}.
Such features have been studied in some detail in the past
with elaborate theoretical approaches, starting from exact
diagonalization\cite{Marel1988,Zaanen1990} 
and more recently within Dynamical Mean-Field Theory 
(DMFT)\cite{Anisimov1997,Lichtenstein1998,Kotliar2006}. 
and are well-documented experimentally 
\cite{Thole1985,Zaanen1990}. 

Another type of satellites appearing in the spectral function 
of electronic materials are due to electrons coupling to
plasmons
\cite{Chang1972,Chang1973,Aryasetiawan2004,Guzzo2011, Casula2012b,Lischner2013,Lemell2015,Borgatti2018}.
Plasmons have been experimentally observed and theoretically investigated 
in materials ranging from 
elementary metals~\cite{Aryasetiawan1995b,Aryasetiawan1996b,Steiner1978},
bronzes\cite{Campagna1975,Chazalviel1977},
oxides\cite{Beatham1980,Aryasetiawan1995,Egdell1999,Christou2000,Kohiki2000,Gatti2007,Borgatti2018,Mudd2014},
in particular ruthenates\cite{Cox1986} and
cuprates\cite{Bozovic1990,Marel2004,Werner2015screen}
as well as in graphene\cite{Xiaoguang2013,Lischner2013,Guzzo2014}.
Plasmonic excitations are relevant and actively utilized in
the design of functional materials\cite{Raveau2005,Cheng2011,Nuraje2012}, such as in
plasmon-mediated photocatalysis\cite{Hou2013,Meng2016}
and sensors\cite{Szunerits2012}.
Plasmons are collective electronic
excitations, which are in general highly non-local in nature. 
They are encoded in the dielectric function
describing the dynamic response of the electronic system as a whole 
to a perturbation.
This response can be mediated by particle-hole excitations as well as by
collective (plasmonic) excitations, which both can give rise
to shake-up satellites in the spectral function. 
For simplicity, below, we will refer to any features beyond a local atomic-like picture, i.e. originating from non-local collective excitations as plasmonic satellites,
and our aim will be to distinguish those from the Hubbard-type satellites
described above.
This differentiation becomes non-trivial when the energy scale of plasmonic
excitations is similar to that of the local Coulomb interactions, which can lead
to both Hubbard and plasmon satellites to appear at similar energies.
Recently, evidence has accumulated that this is the case in a large
number of transition metal oxides \cite{Wang1996,Onari1998,Grosvenor2006,Bansil2007,Gatti2013,Boehnke2016,Nilsson2017,Petocchi2020,Yeh2021}.

In this letter we present a protocol for a quantitative 
\textit{ab initio} identification of Hubbard and plasmonic
contributions in low-energy satellites in real materials. 
Using this protocol, we reinvestigate two prototypical 
$3d$ and $4d$ perovskite transition metal oxides, {\svo} and {\smo},
and determine the Hubbard and plasmonic contributions in the observed low-energy satellites.
Contrary to previous interpretations\cite{Boehnke2016,Nilsson2017,Petocchi2020,Yeh2021}, 
we find both Hubbard and plasmon satellites to be present, albeit with different magnitude.
On the other hand, our findings reconcile seemingly contradictory calculations
within many-body perturbation theory (within the GW approximation) and combined
GW+Dynamical Mean Field Theory (GW+DMFT) in a surprising manner.

{\svo} is a $3d^1$ compound with metallic V $t_{2g}$ states crossing the Fermi level,
forming a typical 3-peak structure in the spectral function, both confirmed from 
experiment\cite{Inoue1995,Rozenberg1996,Sekiyama2004,Takizawa2009,Aizaki2012,Backes2016} and theoretical
calculations\cite{Rozenberg1996,Pavarini2004,Sekiyama2004,Amadon2008,Tomczak2014,Backes2016}. 
The proposed origin of the satellites though has significantly evolved
over the years. Early combined density functional theory and dynamical mean-field theory (DFT+DMFT) 
calculations suggested that the satellites arise from
strong local V-$t_{2g}$ Coulomb interactions in the form of Hubbard 
bands\cite{Rozenberg1996,Liebsch2003,Pavarini2004,Sekiyama2004,Nekrasov2005,Nekrasov2006,Amadon2008}. 
The advent of combined many-body perturbation theory and dynamical
mean field theory ("GW+DMFT") \cite{Biermann2003}, however, made it possible to include both, Hubbard bands and plasmonic features, in the theoretical description, and it was realized  
that in the low energy ($<5$ eV) range features of both types can coexist \cite{Tomczak2012, Tomczak2014}.
Moreover, it was pointed out that the empty V-e$_g$ states that are split
off from the partially filled V-t$_{2g}$ states by the octahedral crystal
field lie in the same energy range as the upper Hubbard band from the
early DFT+DMFT calculations.
Interestingly, many-body perturbation theory alone could also reproduce
the observed satellite features (albeit at slightly shifted energetic
positions) \cite{Gatti2013}, a finding which seemed to be in contradiction 
with the interpretation as Hubbard bands. Along this line, several works
\cite{Gatti2013,Boehnke2016,Nilsson2017,Petocchi2020,Yeh2021}
gave a purely plasmonic interpretation to the lowest energy features
both in the occupied and the unoccupied part of the spectrum.
A new twist appeared when it was realized that oxygen vacancies
contribute spectral weight at the same energy as the satellite in the occupied spectrum\cite{Backes2016}. 
While the importance of oxygen vacancies responsible for part of the
spectral weight in the energy range in question is now widely recognized,
no consensus has been reached so far concerning the origin of the remaining
intrinsic part of the satellite.

We now turn to a brief discussion of the theoretical
description of
the creation of plasmonic features in the spectral function, within DMFT-derived schemes.
As is well-known \cite{Biermann2003,Aryasetiawan2004,Aryasetiawan2006,Casula2012a,Casula2012b,Tomczak2014,Boehnke2016,Reining2018}
electronic screening is a dynamical process, since the response 
of the electronic density in a solid to a perturbation depends on the energy scale of the perturbation.
For a given set of orbitals of interest,
the charge redistribution and thus the screening is energy dependent,
and directly translates into the notion of a frequency-dependent effective screened Coulomb
interaction $U(\omega)$ when higher energy degrees of
freedom are integrated out \cite{Aryasetiawan2006}. 
An approximate form of the effective $U(\omega)$ can be 
obtained for example within the constrained Random-Phase-Approximation (cRPA)~\cite{Aryasetiawan2004,Aryasetiawan2006},
that considers only screening processes outside of a target low-energy subspace.
From $U(\omega)$ two important pieces of physical information can be deduced: First, the value of the static screened interaction
$U(\omega=0)$, determining in an atomic picture the energetic positions of the atomic multiplets,
which in a periodic crystal typically result in non- or weakly dispersive broad satellites, the \textit{Hubbard bands}.
Second, the crossover from the screened to the bare Coulomb interaction at the plasma frequency $\omega_0$ creates satellites from  collective electron excitations
at multiples of $\omega_0$\cite{Casula2012b}.

In oxides with different manifolds of bands (e.g.
corresponding to the $d$- or $p$- states), additional
"subplasmons" corresponding to collective excitations
within specific subspaces of the full Hilbert space
can occur.
For {\svo}, for example, besides the main plasmon (located at $\omega_0 \approx 14.5$~eV) multiple excitations are found
in the dielectric function,
namely around $2.5$~eV and $5$~eV, the former originating from charge-oscillations in the V $t_{2g}$ manifold\cite{Tomczak2012,Gatti2013,Tomczak2014}.
This is precisely the energy scale where
Hubbard satellites have been reported in {\svo}\cite{Rozenberg1996,Pavarini2004,Sekiyama2004,Amadon2008},
indicating that both plasmonic and Hubbard satellite features may exist in this system, and at comparable energies.

\begin{figure}[t]
\includegraphics[width=0.5\textwidth]{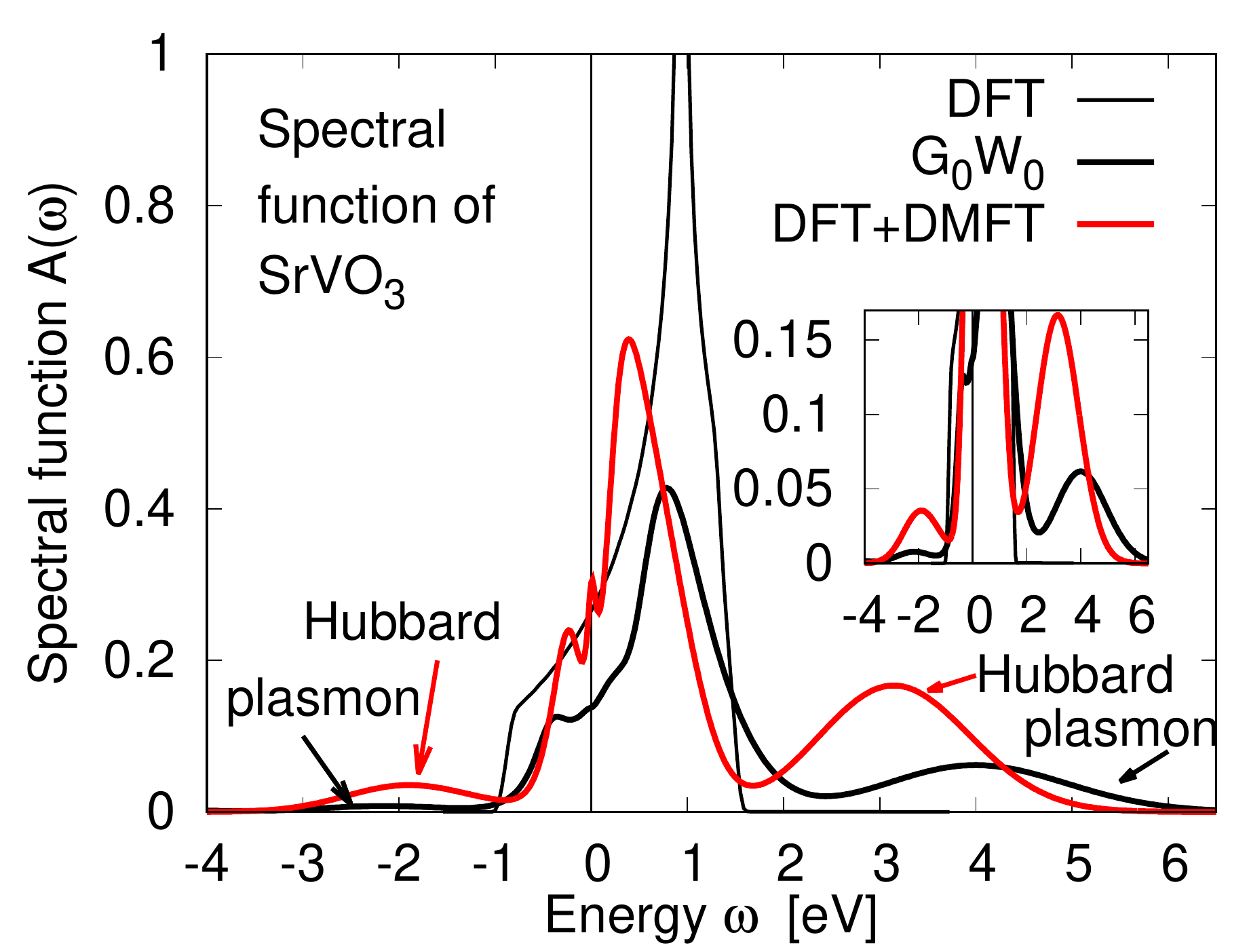} 
\caption{The spectral function of {\svo}, calculated within
Density Functional Theory (DFT), the G$_0$W$_0$ approximation and 
a low-energy model solved in DFT+Dynamical Mean-Field Theory (DFT+DMFT).
The G$_0$W$_0$ approximation introduces corrections due to dynamical 
screening effects and plasmon satellites, while DFT+DMFT describes low-energy
correlations and the emergence of Hubbard bands. 
The inset shows the same data on a smaller scale.
}
\label{fig:srvo3_gw_dmft}
\end{figure}

Different state-of-the-art methods usually obtain only a partial picture of the satellites,
as shown in Fig.~\ref{fig:srvo3_gw_dmft}. Compared to a Density-Functional-Theory (DFT)
calculation, which neither can describe Hubbard or plasmonic satellites,
the consideration of dynamical screening processes within the $GW$ approximation 
introduces plasmonic satellites in the occupied and unoccupied part of the spectrum.
Including the effects of correlations originating from the low-energy part 
$U(\omega=0)$ of the Coulomb interaction but without dynamical screening within DFT+DMFT,
one also observes satellites at very similar energies but now of Hubbard-type origin.
This hints at a possible coexistence of both features in the final spectrum,
but necessitates the use of a method that treats both Hubbard and plasmon contributions
on equal footing, like the combination of $GW$ and DMFT ($GW$+EDMFT)\cite{Biermann2003,Tomczak2012, Tomczak2014,Boehnke2016,Nilsson2017}.
In this method nonlocal correlation and screening processes are accounted
for by the $GW$ approximation, while the local part is obtained from the 
DMFT solution of a local impurity problem subject to the partially screened
interaction $U(\omega)$, which encodes all screening processes beyond 
the low-energy subspace in its frequency dependence.
Since Hubbard satellites originate from the low-energy part
$U(\omega=0)$, and plasmons from dynamical screening, i.e. they emergence in the local
model via the frequency dependence of $U(\omega)$,
we can use this to disentangle their contributions.
Using $GW$+EDMFT in its causal implementation\cite{Backes2022}, we propose the following protocol to
identify and separate out only the plasmonic contributions in the spectral function:
The effective Coulomb interaction $U(\omega)$ can be artificially reduced by a constant shift such that
the static effective Coulomb interaction $U(\omega=0)$ vanishes, but the full frequency dependence 
is retained. This removes contributions from low-energy correlations, \textit{i.e.} the Hubbard
satellites, but fully retains the plasmonic contribution.
(See appendix for a one-orbital proof-of-principle example.)


\begin{figure}[t]
\includegraphics[width=0.5\textwidth]{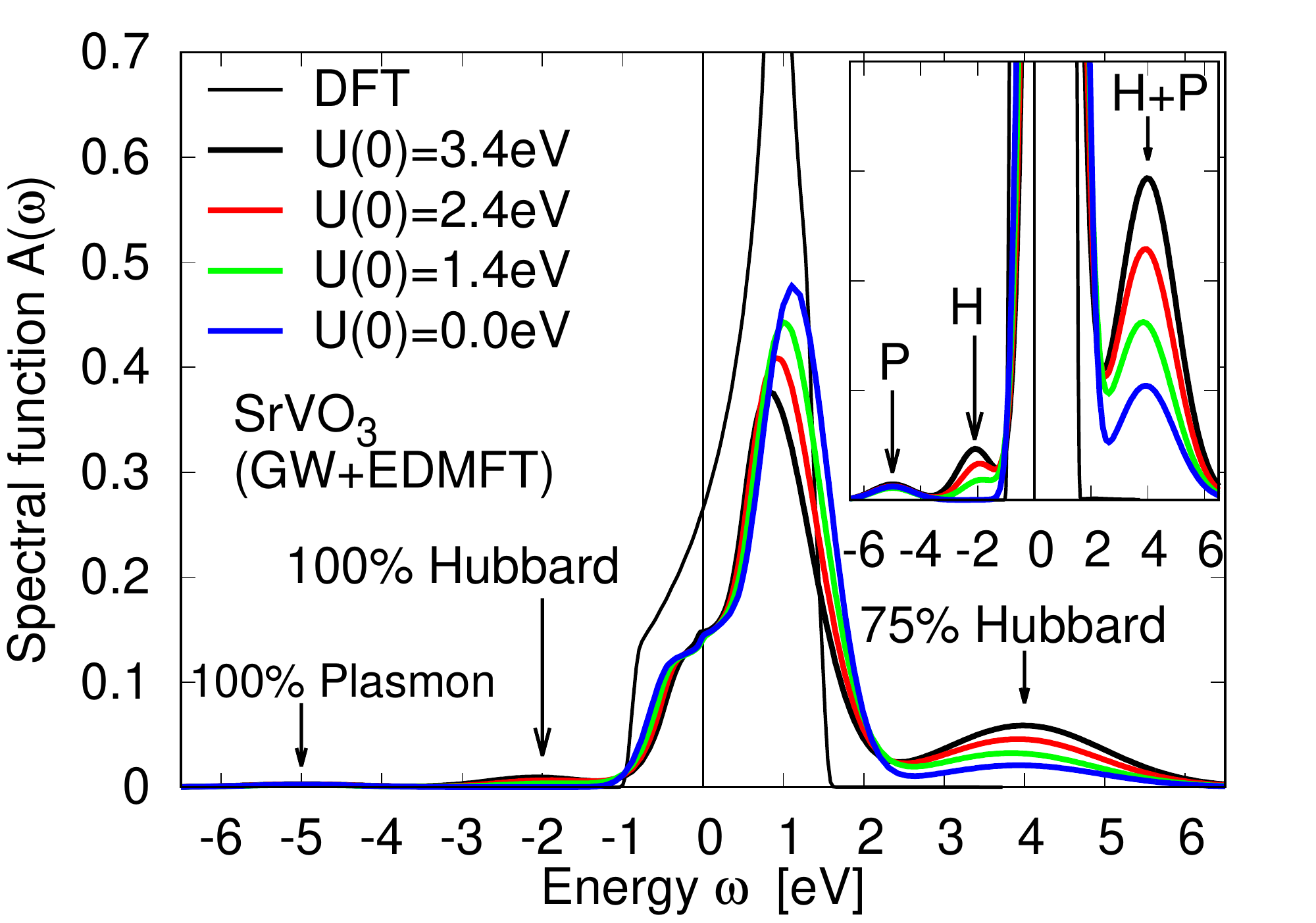} 
\caption{The spectral function of {\svo} for different values of
the screened static interaction $U(0)$ as obtained from GW+DMFT,
including both Hubbard- and plasmonic physics.
The low-energy satellite in the occupied part of the spectrum vanishes for $U(0)=0$~eV,
indicating it is purely composed of a Hubbard satellite.
On the other hand, the upper satellite is composed of $\sim 25/75\%$ 
plasmonic/Hubbard weight.
}
\label{fig:srvo3_umin}
\end{figure}

The resulting spectral function for {\svo} within this scheme
is shown in Fig.~\ref{fig:srvo3_umin}.
Without any artificial reduction of the interaction the result is very similar to previous GW+EDMFT 
calculations\cite{Tomczak2014,Boehnke2016,Nilsson2017}, with a renormalized quasi-particle peak
and a main satellite in the occupied and unoccupied part. Different from DMFT but similar as in GW\cite{Gatti2013,Nakamura2016}
one observes an additional plasmon satellite around $-5$~eV, originating from transitions outside the $t_{2g}$ space\cite{Gatti2013}.
Reducing the static interaction from the \textit{ab initio} value
$U(0)=3.4$~eV to zero,
we observe, besides an expected increase in bandwidth,
a strong reduction of the two satellites closest to the Fermi level, where the lower 
satellite completely vanishes for $U(0)=0$~eV. A small upper 
satellite remains with about $25\%$ of the original weight.
The satellite at $-5$~eV is not affected.
This indicates that the satellite around $-2$~eV
in {\svo} is indeed purely a lower Hubbard band, albeit with an intensity lower than reported in DFT+DMFT.
This in fact agrees with the experimental observation that the lower intrinsic satellite is 
rather small and in general contains significant contributions from oxygen vacancies\cite{Backes2016}.
On the other hand, the remaining satellites around $\pm 5$~eV correspond to the plasmon satellites 
in {\svo} originating from the $5$~eV transition reported in the energy loss function of {\svo}\cite{Tomczak2012,Gatti2013,Tomczak2014}.
The upper satellite is thus composed of Hubbard ($\sim 75\%$) and 
plasmonic ($\sim 25\%$) contributions at similar energies, with the plasmon satellite effectively 'buried'
beneath the dominant Hubbard satellite.

\begin{figure}[t]
\includegraphics[width=0.5\textwidth]{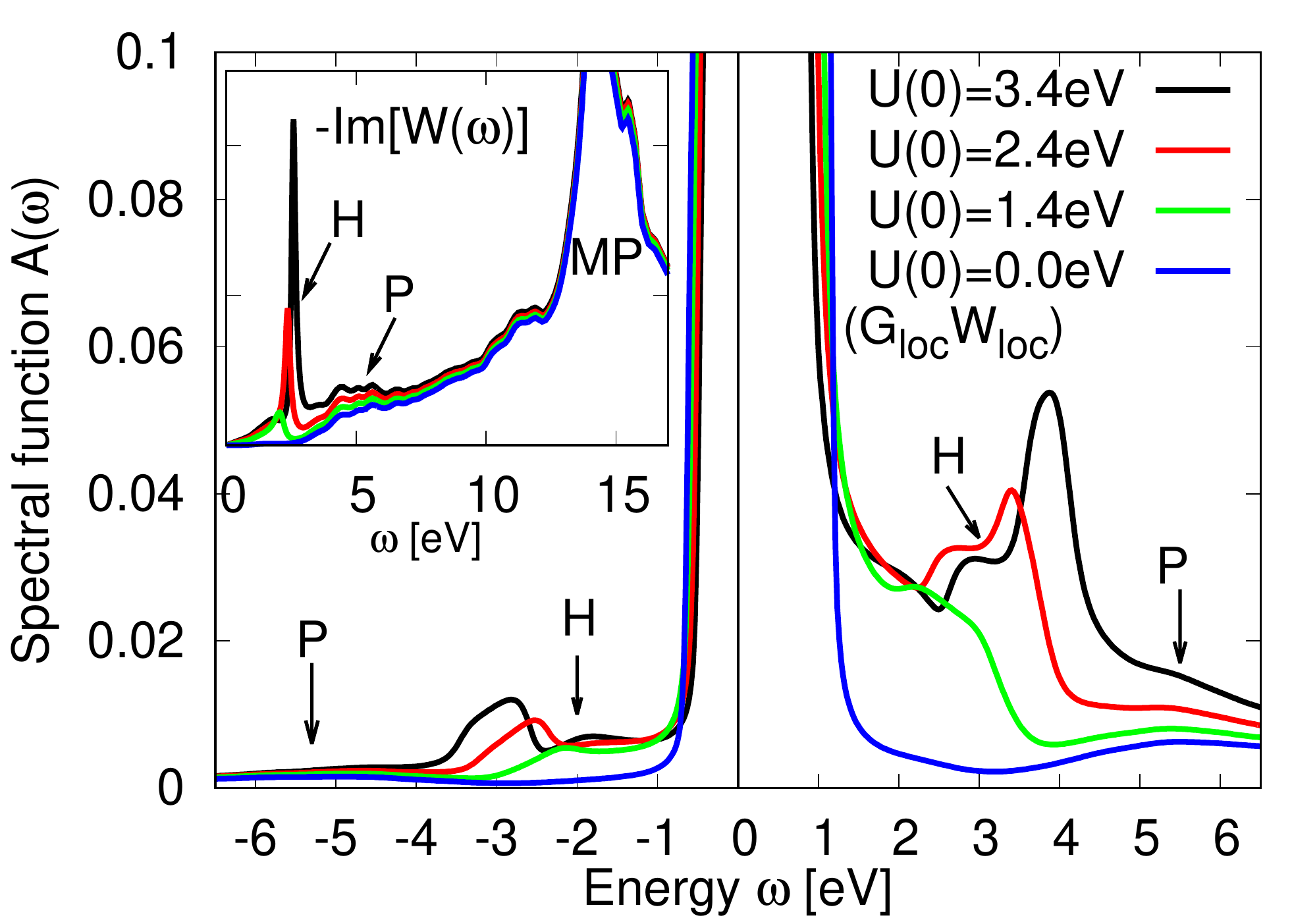} 
\caption{
The t$_{2g}$ spectral function of {\svo} obtained from a local $G_0W_0$ approximation
for different values of the screened interaction $U(0)$ but retaining 
the full frequency dependence. The inset shows the negative imaginary part of the resulting
fully screened interaction $W(\omega)$.
The lower and upper Hubbard (H)-like peaks originate from a local charge oscillation in the $t_{2g}$ orbitals,
corresponding to the peak around $2.5$~eV in $W(\omega)$. As in Fig.\ref{fig:srvo3_umin} these satellites
vanish when the screened static interaction $U(0)$ becomes zero, and only the plasmon
contribution (P) remains. (MP) indicates the main plasmon excitation.
}
\label{fig:srvo3_gwloc}
\end{figure}

Eventually the $GW$+EDMFT spectral function and its satellites are very similar to the $G_0W_0$  result, except
for a slight increase in renormalization (see appendix for a direct comparison),
in contrast to previous results\cite{Boehnke2016,Nilsson2017}, which found a reduction in correlation.
This difference stems from causality violations in the previous computational scheme, as discussed in Ref\cite{Backes2022},
whereas our current scheme does not suffer from this issue.
This agreement between $G_0W_0$ and $GW$+EDMFT not only indicates that the current level of self-consistency is sufficient, but also
that {\svo} is only moderately correlated such that $G_0W_0$ is able to capture most of the relevant physics.
Therefore, the interpretation of the low-energy satellites as Hubbard satellites in {\svo}
raises the question about the true nature of the $G_0W_0$ low-energy satellites.
As they originate from charge excitations in the vanadium $t_{2g}$ manifold\cite{Gatti2013,Tomczak2014},
we apply a similar local $G_0W_0$ scheme to disentangle possible collective non-local charge excitations
(plasmons) from local Hubbard-like physics.
In Fig.~\ref{fig:srvo3_gwloc} we show the resulting spectral function $A(\omega)$ and screened
interaction $W(\omega)$ for {\svo} within a local low-energy $G_0W_0$ scheme. In this scheme 
the local 'bare' V $t_{2g}$ interaction $U(\omega)$ is screened by only considering local transitions
in the $t_{2g}$ space, and the resulting $W(\omega)$ is convoluted with the local non-interacting $t_{2g}$
Green's function to obtain the effective self-energy (i.e., the impurity model is solved within the $G_0W_0$ approximation).
The resulting $W(\omega)$ and spectral function almost perfectly reproduces the full $G_0W_0$ calculation, besides
an overestimation of the energetic position of the $t_{2g}$ derived peak in $W(\omega)$ around $3$~eV,
which leads to an overestimation of the satellite position. As the calculation has been performed on the real frequency axis,
more pronounced structures are visible and not smeared out by the analytic continuation procedure.
This result indicates that the low-energy peaks in $G_0W_0$ can be explained by only considering local
charge excitations and a local Coulomb interaction. Similarly as for the $GW$+EDMFT result,
the peaks vanish as the static screened interaction is reduced, confirming their local 'Hubbard'-like nature.
Even though $G_0W_0$ as a perturbative approach cannot access strong electronic correlations, 
the corresponding atomic multiplet excitations are effectively encoded in $W(\omega)$ via the RPA approximation and give 
rise to satellites representing the Hubbard satellites obtained in non-perturbative methods.
Thus, this result confirms that the low-energy satellites in {\svo} do not originate from non-local
collective excitations but arise purely from local Hubbard-like charge excitations given by the static
local interaction $U(\omega=0)$. Plasmon satellites are only found at energies (and beyond) $\pm 5$~eV.

\begin{figure}[t]
\includegraphics[width=0.5\textwidth]{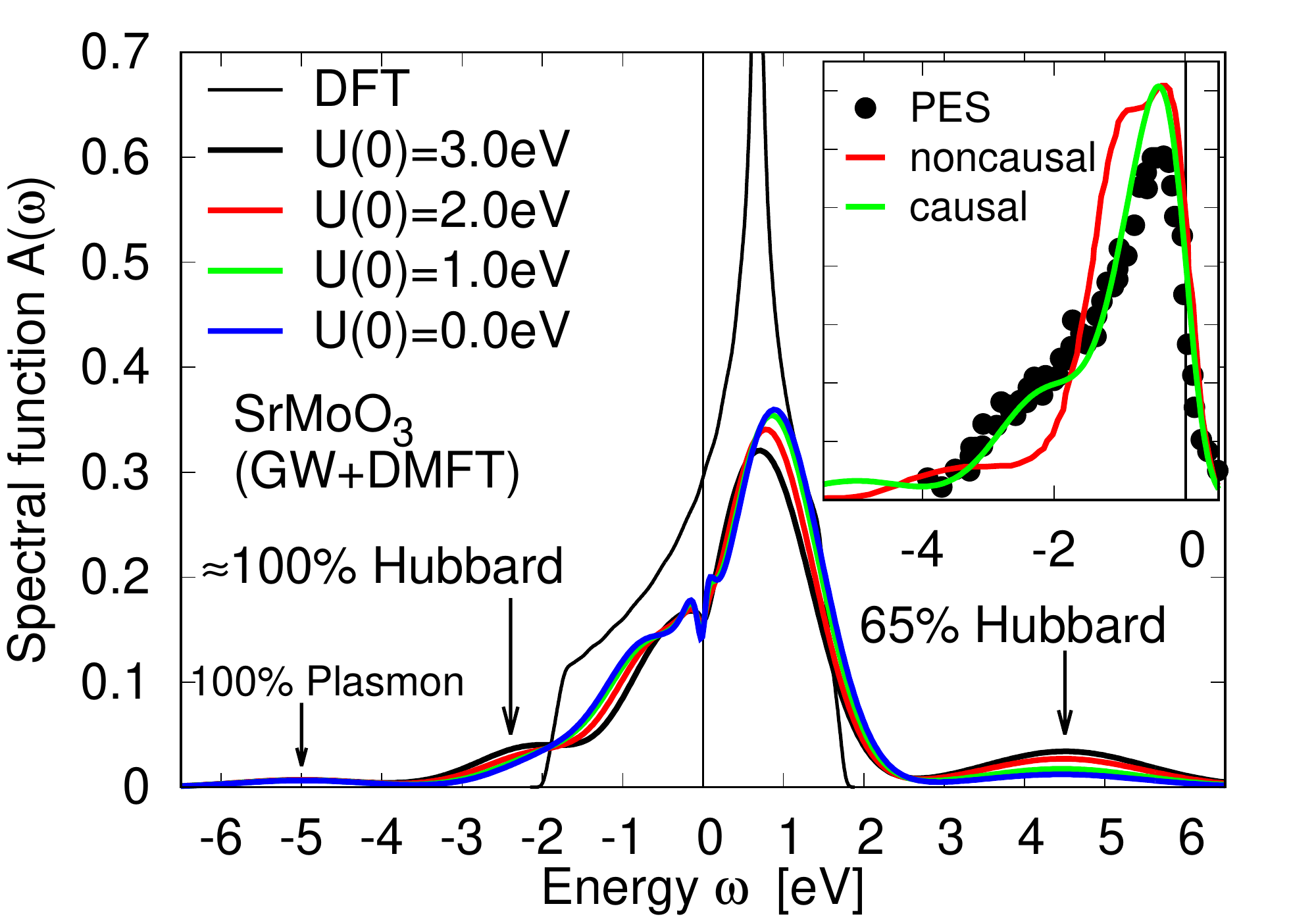} 
\caption{The spectral function of {\smo} for different values of
the screened static interaction $U(0)$ as obtained from $GW$+EDMFT.
The low-energy satellite around $-2.5$~eV is mostly composed of a Hubbard satellite,
while the unoccupied satellite is about $35\%$ plasmonic and $65\%$ Hubbard type origin.
The inset shows the same spectral function at $U=3$~eV, compared to a noncausal implementation (taken from~\cite{Nilsson2017})
and photoemission data (taken from~\cite{Ali2019}).
}
\label{fig:srmoo3_umin}
\end{figure}

We apply the same method to the closely related $4d^2$ material {\smo}, which is 
isostructural to {\svo}. Due to the more extended nature of the $4d$ orbitals,
low-energy electronic correlations are weaker and the experimentally observed
satellites have been proposed to be of purely plasmonic origin\cite{Wadati2014,Radetinac2016,Nilsson2017,Ali2019}.
The resulting spectral function for different values of the static interaction
is shown in Fig.~\ref{fig:srmoo3_umin}. 
As in previous reports we obtain a much broader quasiparticle peak than in
{\svo}, with a lower shoulder-like feature around $2.5$~eV and an upper satellite. Upon reducing the static interaction,
we see a similar trend as in {\svo}, but less pronounced. For $U(0)=0$ the occupied shoulder 
seems to completely merge with the quasiparticle peak, corresponding to a Hubbard satellite,
but we found the analytic continuation procedure to be less reliable in that area.
The existence of a small lower Hubbard satellite
does not contradict with a previous DFT+DMFT study\cite{Wadati2014}, 
which found no pronounced Hubbard satellite but still a significant amount of spectral weight shifted to
lower energies around $-2.5$~eV, very similar to our results.
The unoccupied satellite is significantly reduced, retaining about $35\%$ of its weight as a plasmon satellite.
Similar as in {\svo}, we find that {\smo} shows pronounced plasmon satellites around $\pm 5$~eV.
Most noteworthy, the causal implementation of $GW$+EDMFT used in this manuscript is able to accurately reproduce 
the observed experimental spectral function from photoemission experiments\cite{Ali2019},
demonstrating that this approach is capable of capturing the relevant physics in this material and thus
strengthening our interpretation of the spectral features as Hubbard satellites.
The noncausal formulation\cite{Nilsson2017} shows a significantly lower intensity for the Hubbard-like satellite
and increased bandwidth of the V $t_{2g}$ quasi-particle dispersive states, indicating that the noncausal variant
underestimates the electronic correlation strength also in {\smo}, in line with previous observations\cite{Backes2022,Chen2022}.

In summary, we have revisited the decade-old problem of the nature of
spectral satellites in ternary transition metal oxides, 
and proposed a theoretical \textit{ab initio} method to distinguish 
plasmonic satellites, which emerge from collective electronic excitations,
from Hubbard-type satellites, resulting from a strong local Coulomb interaction.
For the prototypical transition metal oxide {\svo} 
we show that the occupied low-energy satellite is purely
composed of Hubbard-type incoherent weight, while in the unoccupied satellite
both Hubbard and plasmonic contributions coexist at similar energies. 
In the weaker correlated $4d^2$ {\smo} we observe a similar but less pronounced picture of a 
Hubbard satellite in the occupied
part, and both plasmon and Hubbard satellites at similar energies in the unoccupied part. 
These observation call for a reinvestigation of similar and other correlated materials
and their satellite features, both theoretically and experimentally,
as they are in particular relevant for plasmon-mediated applications in functional materials,
where precise knowledge of the intensity and energy of plasmonic excitations is needed.
The scheme that we have developed can be applied to a large class of materials,
and can aid the development of such applications by theoretically quantifying
the plasmonic and Hubbard-type contributions in the spectral satellites.

While this work was being prepared for publication,
a new joint experimental-theoretical work appeared on SrVO$_3$\cite{Gloter2023}.
In that work, a purely plasmonic origin of the satellites
is advocated. However, we believe this apparent contradiction
also to be resolved by our work, since we show that the
corresponding features stem indeed from the dielectric
function, but are nevertheless of multiplet-like origin rather
than long-range collective excitations, see discussion above.

\acknowledgments 

The authors gratefully acknowledge fruitful discussions with F. Aryasetiawan, M. Gatti, A. Lichtenstein, L. Reining and G. Sawatzky.
This work was supported by a Consolidator Grant
of the European Research Council (Project CorrelMat-617196)
and GENCI/IDRIS Orsay under project A0130901393.

\bibliographystyle{apsrev4-2}
\bibliography{materials,refs}

\begin{thebibliography}{75}%
\makeatletter
\providecommand \@ifxundefined [1]{%
 \@ifx{#1\undefined}
}%
\providecommand \@ifnum [1]{%
 \ifnum #1\expandafter \@firstoftwo
 \else \expandafter \@secondoftwo
 \fi
}%
\providecommand \@ifx [1]{%
 \ifx #1\expandafter \@firstoftwo
 \else \expandafter \@secondoftwo
 \fi
}%
\providecommand \natexlab [1]{#1}%
\providecommand \enquote  [1]{``#1''}%
\providecommand \bibnamefont  [1]{#1}%
\providecommand \bibfnamefont [1]{#1}%
\providecommand \citenamefont [1]{#1}%
\providecommand \href@noop [0]{\@secondoftwo}%
\providecommand \href [0]{\begingroup \@sanitize@url \@href}%
\providecommand \@href[1]{\@@startlink{#1}\@@href}%
\providecommand \@@href[1]{\endgroup#1\@@endlink}%
\providecommand \@sanitize@url [0]{\catcode `\\12\catcode `\$12\catcode
  `\&12\catcode `\#12\catcode `\^12\catcode `\_12\catcode `\%12\relax}%
\providecommand \@@startlink[1]{}%
\providecommand \@@endlink[0]{}%
\providecommand \url  [0]{\begingroup\@sanitize@url \@url }%
\providecommand \@url [1]{\endgroup\@href {#1}{\urlprefix }}%
\providecommand \urlprefix  [0]{URL }%
\providecommand \Eprint [0]{\href }%
\providecommand \doibase [0]{https://doi.org/}%
\providecommand \selectlanguage [0]{\@gobble}%
\providecommand \bibinfo  [0]{\@secondoftwo}%
\providecommand \bibfield  [0]{\@secondoftwo}%
\providecommand \translation [1]{[#1]}%
\providecommand \BibitemOpen [0]{}%
\providecommand \bibitemStop [0]{}%
\providecommand \bibitemNoStop [0]{.\EOS\space}%
\providecommand \EOS [0]{\spacefactor3000\relax}%
\providecommand \BibitemShut  [1]{\csname bibitem#1\endcsname}%
\let\auto@bib@innerbib\@empty
\bibitem [{\citenamefont {Damascelli}(2004)}]{Damascelli2004}%
  \BibitemOpen
  \bibfield  {author} {\bibinfo {author} {\bibfnamefont {A.}~\bibnamefont
  {Damascelli}},\ }\bibfield  {journal} {\bibinfo  {journal} {Physica Scripta
  Volume T}\ }\textbf {\bibinfo {volume} {109}},\ \href
  {https://doi.org/10.1238/Physica.Topical.109a00061}
  {10.1238/Physica.Topical.109a00061} (\bibinfo {year} {2004})\BibitemShut
  {NoStop}%
\bibitem [{\citenamefont {Sobota}\ \emph {et~al.}(2021)\citenamefont {Sobota},
  \citenamefont {He},\ and\ \citenamefont {Shen}}]{Sobota2021}%
  \BibitemOpen
  \bibfield  {author} {\bibinfo {author} {\bibfnamefont {J.~A.}\ \bibnamefont
  {Sobota}}, \bibinfo {author} {\bibfnamefont {Y.}~\bibnamefont {He}},\ and\
  \bibinfo {author} {\bibfnamefont {Z.-X.}\ \bibnamefont {Shen}},\ }\href
  {https://doi.org/10.1103/RevModPhys.93.025006} {\bibfield  {journal}
  {\bibinfo  {journal} {Rev. Mod. Phys.}\ }\textbf {\bibinfo {volume} {93}},\
  \bibinfo {pages} {025006} (\bibinfo {year} {2021})}\BibitemShut {NoStop}%
\bibitem [{\citenamefont {Zhang}\ \emph {et~al.}(2022)\citenamefont {Zhang},
  \citenamefont {Pincelli}, \citenamefont {Jozwiak}, \citenamefont {Kondo},
  \citenamefont {Ernstorfer}, \citenamefont {Sato},\ and\ \citenamefont
  {Zhou}}]{Zhang2022}%
  \BibitemOpen
  \bibfield  {author} {\bibinfo {author} {\bibfnamefont {H.}~\bibnamefont
  {Zhang}}, \bibinfo {author} {\bibfnamefont {T.}~\bibnamefont {Pincelli}},
  \bibinfo {author} {\bibfnamefont {C.}~\bibnamefont {Jozwiak}}, \bibinfo
  {author} {\bibfnamefont {T.}~\bibnamefont {Kondo}}, \bibinfo {author}
  {\bibfnamefont {R.}~\bibnamefont {Ernstorfer}}, \bibinfo {author}
  {\bibfnamefont {T.}~\bibnamefont {Sato}},\ and\ \bibinfo {author}
  {\bibfnamefont {S.}~\bibnamefont {Zhou}},\ }\bibfield  {journal} {\bibinfo
  {journal} {Nature Reviews Methods Primers}\ }\textbf {\bibinfo {volume}
  {2}},\ \href {https://doi.org/10.1038/s43586-022-00133-7}
  {10.1038/s43586-022-00133-7} (\bibinfo {year} {2022})\BibitemShut {NoStop}%
\bibitem [{\citenamefont {Hubbard}(1963)}]{Hubbard1963}%
  \BibitemOpen
  \bibfield  {author} {\bibinfo {author} {\bibfnamefont {J.}~\bibnamefont
  {Hubbard}},\ }\href {https://doi.org/10.1098/rspa.1963.0204} {\bibfield
  {journal} {\bibinfo  {journal} {Proceedings of the Royal Society of London A:
  Mathematical, Physical and Engineering Sciences}\ }\textbf {\bibinfo {volume}
  {276}},\ \bibinfo {pages} {238} (\bibinfo {year} {1963})},\ \Eprint
  {https://arxiv.org/abs/http://rspa.royalsocietypublishing.org/content/276/1365/238.full.pdf}
  {http://rspa.royalsocietypublishing.org/content/276/1365/238.full.pdf}
  \BibitemShut {NoStop}%
\bibitem [{\citenamefont {van~der Marel}\ and\ \citenamefont
  {Sawatzky}(1988)}]{Marel1988}%
  \BibitemOpen
  \bibfield  {author} {\bibinfo {author} {\bibfnamefont {D.}~\bibnamefont
  {van~der Marel}}\ and\ \bibinfo {author} {\bibfnamefont {G.~A.}\ \bibnamefont
  {Sawatzky}},\ }\href {https://doi.org/10.1103/PhysRevB.37.10674} {\bibfield
  {journal} {\bibinfo  {journal} {Phys. Rev. B}\ }\textbf {\bibinfo {volume}
  {37}},\ \bibinfo {pages} {10674} (\bibinfo {year} {1988})}\BibitemShut
  {NoStop}%
\bibitem [{\citenamefont {Zaanen}\ and\ \citenamefont
  {Sawatzky}(1990)}]{Zaanen1990}%
  \BibitemOpen
  \bibfield  {author} {\bibinfo {author} {\bibfnamefont {J.}~\bibnamefont
  {Zaanen}}\ and\ \bibinfo {author} {\bibfnamefont {G.~A.}\ \bibnamefont
  {Sawatzky}},\ }\href {https://doi.org/10.1143/PTP.101.231} {\bibfield
  {journal} {\bibinfo  {journal} {Progress of Theoretical Physics Supplement}\
  }\textbf {\bibinfo {volume} {101}},\ \bibinfo {pages} {231} (\bibinfo {year}
  {1990})},\ \Eprint
  {https://arxiv.org/abs/https://academic.oup.com/ptps/article-pdf/doi/10.1143/PTP.101.231/5450560/101-231.pdf}
  {https://academic.oup.com/ptps/article-pdf/doi/10.1143/PTP.101.231/5450560/101-231.pdf}
  \BibitemShut {NoStop}%
\bibitem [{\citenamefont {Anisimov}\ \emph {et~al.}(1997)\citenamefont
  {Anisimov}, \citenamefont {Poteryaev}, \citenamefont {Korotin}, \citenamefont
  {Anokhin},\ and\ \citenamefont {Kotliar}}]{Anisimov1997}%
  \BibitemOpen
  \bibfield  {author} {\bibinfo {author} {\bibfnamefont {V.~I.}\ \bibnamefont
  {Anisimov}}, \bibinfo {author} {\bibfnamefont {A.~I.}\ \bibnamefont
  {Poteryaev}}, \bibinfo {author} {\bibfnamefont {M.~A.}\ \bibnamefont
  {Korotin}}, \bibinfo {author} {\bibfnamefont {A.~O.}\ \bibnamefont
  {Anokhin}},\ and\ \bibinfo {author} {\bibfnamefont {G.}~\bibnamefont
  {Kotliar}},\ }\href {http://stacks.iop.org/0953-8984/9/i=35/a=010} {\bibfield
   {journal} {\bibinfo  {journal} {Journal of Physics: Condensed Matter}\
  }\textbf {\bibinfo {volume} {9}},\ \bibinfo {pages} {7359} (\bibinfo {year}
  {1997})}\BibitemShut {NoStop}%
\bibitem [{\citenamefont {Lichtenstein}\ and\ \citenamefont
  {Katsnelson}(1998)}]{Lichtenstein1998}%
  \BibitemOpen
  \bibfield  {author} {\bibinfo {author} {\bibfnamefont {A.~I.}\ \bibnamefont
  {Lichtenstein}}\ and\ \bibinfo {author} {\bibfnamefont {M.~I.}\ \bibnamefont
  {Katsnelson}},\ }\href {https://doi.org/10.1103/PhysRevB.57.6884} {\bibfield
  {journal} {\bibinfo  {journal} {Phys. Rev. B}\ }\textbf {\bibinfo {volume}
  {57}},\ \bibinfo {pages} {6884} (\bibinfo {year} {1998})}\BibitemShut
  {NoStop}%
\bibitem [{\citenamefont {Kotliar}\ \emph {et~al.}(2006)\citenamefont
  {Kotliar}, \citenamefont {Savrasov}, \citenamefont {Haule}, \citenamefont
  {Oudovenko}, \citenamefont {Parcollet},\ and\ \citenamefont
  {Marianetti}}]{Kotliar2006}%
  \BibitemOpen
  \bibfield  {author} {\bibinfo {author} {\bibfnamefont {G.}~\bibnamefont
  {Kotliar}}, \bibinfo {author} {\bibfnamefont {S.~Y.}\ \bibnamefont
  {Savrasov}}, \bibinfo {author} {\bibfnamefont {K.}~\bibnamefont {Haule}},
  \bibinfo {author} {\bibfnamefont {V.~S.}\ \bibnamefont {Oudovenko}}, \bibinfo
  {author} {\bibfnamefont {O.}~\bibnamefont {Parcollet}},\ and\ \bibinfo
  {author} {\bibfnamefont {C.~A.}\ \bibnamefont {Marianetti}},\ }\href
  {https://doi.org/10.1103/RevModPhys.78.865} {\bibfield  {journal} {\bibinfo
  {journal} {Rev. Mod. Phys.}\ }\textbf {\bibinfo {volume} {78}},\ \bibinfo
  {pages} {865} (\bibinfo {year} {2006})}\BibitemShut {NoStop}%
\bibitem [{\citenamefont {Thole}\ \emph {et~al.}(1985)\citenamefont {Thole},
  \citenamefont {van~der Laan}, \citenamefont {Fuggle}, \citenamefont
  {Sawatzky}, \citenamefont {Karnatak},\ and\ \citenamefont
  {Esteva}}]{Thole1985}%
  \BibitemOpen
  \bibfield  {author} {\bibinfo {author} {\bibfnamefont {B.~T.}\ \bibnamefont
  {Thole}}, \bibinfo {author} {\bibfnamefont {G.}~\bibnamefont {van~der Laan}},
  \bibinfo {author} {\bibfnamefont {J.~C.}\ \bibnamefont {Fuggle}}, \bibinfo
  {author} {\bibfnamefont {G.~A.}\ \bibnamefont {Sawatzky}}, \bibinfo {author}
  {\bibfnamefont {R.~C.}\ \bibnamefont {Karnatak}},\ and\ \bibinfo {author}
  {\bibfnamefont {J.-M.}\ \bibnamefont {Esteva}},\ }\href
  {https://doi.org/10.1103/PhysRevB.32.5107} {\bibfield  {journal} {\bibinfo
  {journal} {Phys. Rev. B}\ }\textbf {\bibinfo {volume} {32}},\ \bibinfo
  {pages} {5107} (\bibinfo {year} {1985})}\BibitemShut {NoStop}%
\bibitem [{\citenamefont {Chang}\ and\ \citenamefont
  {Langreth}(1972)}]{Chang1972}%
  \BibitemOpen
  \bibfield  {author} {\bibinfo {author} {\bibfnamefont {J.-J.}\ \bibnamefont
  {Chang}}\ and\ \bibinfo {author} {\bibfnamefont {D.~C.}\ \bibnamefont
  {Langreth}},\ }\href {https://doi.org/10.1103/PhysRevB.5.3512} {\bibfield
  {journal} {\bibinfo  {journal} {Phys. Rev. B}\ }\textbf {\bibinfo {volume}
  {5}},\ \bibinfo {pages} {3512} (\bibinfo {year} {1972})}\BibitemShut
  {NoStop}%
\bibitem [{\citenamefont {Chang}\ and\ \citenamefont
  {Langreth}(1973)}]{Chang1973}%
  \BibitemOpen
  \bibfield  {author} {\bibinfo {author} {\bibfnamefont {J.-J.}\ \bibnamefont
  {Chang}}\ and\ \bibinfo {author} {\bibfnamefont {D.~C.}\ \bibnamefont
  {Langreth}},\ }\href {https://doi.org/10.1103/PhysRevB.8.4638} {\bibfield
  {journal} {\bibinfo  {journal} {Phys. Rev. B}\ }\textbf {\bibinfo {volume}
  {8}},\ \bibinfo {pages} {4638} (\bibinfo {year} {1973})}\BibitemShut
  {NoStop}%
\bibitem [{\citenamefont {Aryasetiawan}\ \emph {et~al.}(2004)\citenamefont
  {Aryasetiawan}, \citenamefont {Imada}, \citenamefont {Georges}, \citenamefont
  {Kotliar}, \citenamefont {Biermann},\ and\ \citenamefont
  {Lichtenstein}}]{Aryasetiawan2004}%
  \BibitemOpen
  \bibfield  {author} {\bibinfo {author} {\bibfnamefont {F.}~\bibnamefont
  {Aryasetiawan}}, \bibinfo {author} {\bibfnamefont {M.}~\bibnamefont {Imada}},
  \bibinfo {author} {\bibfnamefont {A.}~\bibnamefont {Georges}}, \bibinfo
  {author} {\bibfnamefont {G.}~\bibnamefont {Kotliar}}, \bibinfo {author}
  {\bibfnamefont {S.}~\bibnamefont {Biermann}},\ and\ \bibinfo {author}
  {\bibfnamefont {A.~I.}\ \bibnamefont {Lichtenstein}},\ }\href
  {https://doi.org/10.1103/PhysRevB.70.195104} {\bibfield  {journal} {\bibinfo
  {journal} {Phys. Rev. B}\ }\textbf {\bibinfo {volume} {70}},\ \bibinfo
  {pages} {195104} (\bibinfo {year} {2004})}\BibitemShut {NoStop}%
\bibitem [{\citenamefont {Guzzo}\ \emph {et~al.}(2011)\citenamefont {Guzzo},
  \citenamefont {Lani}, \citenamefont {Sottile}, \citenamefont {Romaniello},
  \citenamefont {Gatti}, \citenamefont {Kas}, \citenamefont {Rehr},
  \citenamefont {Silly}, \citenamefont {Sirotti},\ and\ \citenamefont
  {Reining}}]{Guzzo2011}%
  \BibitemOpen
  \bibfield  {author} {\bibinfo {author} {\bibfnamefont {M.}~\bibnamefont
  {Guzzo}}, \bibinfo {author} {\bibfnamefont {G.}~\bibnamefont {Lani}},
  \bibinfo {author} {\bibfnamefont {F.}~\bibnamefont {Sottile}}, \bibinfo
  {author} {\bibfnamefont {P.}~\bibnamefont {Romaniello}}, \bibinfo {author}
  {\bibfnamefont {M.}~\bibnamefont {Gatti}}, \bibinfo {author} {\bibfnamefont
  {J.~J.}\ \bibnamefont {Kas}}, \bibinfo {author} {\bibfnamefont {J.~J.}\
  \bibnamefont {Rehr}}, \bibinfo {author} {\bibfnamefont {M.~G.}\ \bibnamefont
  {Silly}}, \bibinfo {author} {\bibfnamefont {F.}~\bibnamefont {Sirotti}},\
  and\ \bibinfo {author} {\bibfnamefont {L.}~\bibnamefont {Reining}},\ }\href
  {https://doi.org/10.1103/PhysRevLett.107.166401} {\bibfield  {journal}
  {\bibinfo  {journal} {Phys. Rev. Lett.}\ }\textbf {\bibinfo {volume} {107}},\
  \bibinfo {pages} {166401} (\bibinfo {year} {2011})}\BibitemShut {NoStop}%
\bibitem [{\citenamefont {Casula}\ \emph
  {et~al.}(2012{\natexlab{a}})\citenamefont {Casula}, \citenamefont {Werner},
  \citenamefont {Vaugier}, \citenamefont {Aryasetiawan}, \citenamefont
  {Miyake}, \citenamefont {Millis},\ and\ \citenamefont
  {Biermann}}]{Casula2012b}%
  \BibitemOpen
  \bibfield  {author} {\bibinfo {author} {\bibfnamefont {M.}~\bibnamefont
  {Casula}}, \bibinfo {author} {\bibfnamefont {P.}~\bibnamefont {Werner}},
  \bibinfo {author} {\bibfnamefont {L.}~\bibnamefont {Vaugier}}, \bibinfo
  {author} {\bibfnamefont {F.}~\bibnamefont {Aryasetiawan}}, \bibinfo {author}
  {\bibfnamefont {T.}~\bibnamefont {Miyake}}, \bibinfo {author} {\bibfnamefont
  {A.~J.}\ \bibnamefont {Millis}},\ and\ \bibinfo {author} {\bibfnamefont
  {S.}~\bibnamefont {Biermann}},\ }\href
  {https://doi.org/10.1103/PhysRevLett.109.126408} {\bibfield  {journal}
  {\bibinfo  {journal} {Phys. Rev. Lett.}\ }\textbf {\bibinfo {volume} {109}},\
  \bibinfo {pages} {126408} (\bibinfo {year} {2012}{\natexlab{a}})}\BibitemShut
  {NoStop}%
\bibitem [{\citenamefont {Lischner}\ \emph {et~al.}(2013)\citenamefont
  {Lischner}, \citenamefont {Vigil-Fowler},\ and\ \citenamefont
  {Louie}}]{Lischner2013}%
  \BibitemOpen
  \bibfield  {author} {\bibinfo {author} {\bibfnamefont {J.}~\bibnamefont
  {Lischner}}, \bibinfo {author} {\bibfnamefont {D.}~\bibnamefont
  {Vigil-Fowler}},\ and\ \bibinfo {author} {\bibfnamefont {S.~G.}\ \bibnamefont
  {Louie}},\ }\href {https://doi.org/10.1103/PhysRevLett.110.146801} {\bibfield
   {journal} {\bibinfo  {journal} {Phys. Rev. Lett.}\ }\textbf {\bibinfo
  {volume} {110}},\ \bibinfo {pages} {146801} (\bibinfo {year}
  {2013})}\BibitemShut {NoStop}%
\bibitem [{\citenamefont {Lemell}\ \emph {et~al.}(2015)\citenamefont {Lemell},
  \citenamefont {Neppl}, \citenamefont {Wachter}, \citenamefont
  {T\ifmmode~\mbox{\H{o}}\else \H{o}\fi{}k\'esi}, \citenamefont {Ernstorfer},
  \citenamefont {Feulner}, \citenamefont {Kienberger},\ and\ \citenamefont
  {Burgd\"orfer}}]{Lemell2015}%
  \BibitemOpen
  \bibfield  {author} {\bibinfo {author} {\bibfnamefont {C.}~\bibnamefont
  {Lemell}}, \bibinfo {author} {\bibfnamefont {S.}~\bibnamefont {Neppl}},
  \bibinfo {author} {\bibfnamefont {G.}~\bibnamefont {Wachter}}, \bibinfo
  {author} {\bibfnamefont {K.}~\bibnamefont {T\ifmmode~\mbox{\H{o}}\else
  \H{o}\fi{}k\'esi}}, \bibinfo {author} {\bibfnamefont {R.}~\bibnamefont
  {Ernstorfer}}, \bibinfo {author} {\bibfnamefont {P.}~\bibnamefont {Feulner}},
  \bibinfo {author} {\bibfnamefont {R.}~\bibnamefont {Kienberger}},\ and\
  \bibinfo {author} {\bibfnamefont {J.}~\bibnamefont {Burgd\"orfer}},\ }\href
  {https://doi.org/10.1103/PhysRevB.91.241101} {\bibfield  {journal} {\bibinfo
  {journal} {Phys. Rev. B}\ }\textbf {\bibinfo {volume} {91}},\ \bibinfo
  {pages} {241101} (\bibinfo {year} {2015})}\BibitemShut {NoStop}%
\bibitem [{\citenamefont {Borgatti}\ \emph {et~al.}(2018)\citenamefont
  {Borgatti}, \citenamefont {Berger}, \citenamefont {C\'eolin}, \citenamefont
  {Zhou}, \citenamefont {Kas}, \citenamefont {Guzzo}, \citenamefont
  {McConville}, \citenamefont {Offi}, \citenamefont {Panaccione}, \citenamefont
  {Regoutz}, \citenamefont {Payne}, \citenamefont {Rueff}, \citenamefont
  {Bierwagen}, \citenamefont {White}, \citenamefont {Speck}, \citenamefont
  {Gatti},\ and\ \citenamefont {Egdell}}]{Borgatti2018}%
  \BibitemOpen
  \bibfield  {author} {\bibinfo {author} {\bibfnamefont {F.}~\bibnamefont
  {Borgatti}}, \bibinfo {author} {\bibfnamefont {J.~A.}\ \bibnamefont
  {Berger}}, \bibinfo {author} {\bibfnamefont {D.}~\bibnamefont {C\'eolin}},
  \bibinfo {author} {\bibfnamefont {J.~S.}\ \bibnamefont {Zhou}}, \bibinfo
  {author} {\bibfnamefont {J.~J.}\ \bibnamefont {Kas}}, \bibinfo {author}
  {\bibfnamefont {M.}~\bibnamefont {Guzzo}}, \bibinfo {author} {\bibfnamefont
  {C.~F.}\ \bibnamefont {McConville}}, \bibinfo {author} {\bibfnamefont
  {F.}~\bibnamefont {Offi}}, \bibinfo {author} {\bibfnamefont {G.}~\bibnamefont
  {Panaccione}}, \bibinfo {author} {\bibfnamefont {A.}~\bibnamefont {Regoutz}},
  \bibinfo {author} {\bibfnamefont {D.~J.}\ \bibnamefont {Payne}}, \bibinfo
  {author} {\bibfnamefont {J.-P.}\ \bibnamefont {Rueff}}, \bibinfo {author}
  {\bibfnamefont {O.}~\bibnamefont {Bierwagen}}, \bibinfo {author}
  {\bibfnamefont {M.~E.}\ \bibnamefont {White}}, \bibinfo {author}
  {\bibfnamefont {J.~S.}\ \bibnamefont {Speck}}, \bibinfo {author}
  {\bibfnamefont {M.}~\bibnamefont {Gatti}},\ and\ \bibinfo {author}
  {\bibfnamefont {R.~G.}\ \bibnamefont {Egdell}},\ }\href
  {https://doi.org/10.1103/PhysRevB.97.155102} {\bibfield  {journal} {\bibinfo
  {journal} {Phys. Rev. B}\ }\textbf {\bibinfo {volume} {97}},\ \bibinfo
  {pages} {155102} (\bibinfo {year} {2018})}\BibitemShut {NoStop}%
\bibitem [{\citenamefont {Karlsson}\ and\ \citenamefont
  {Aryasetiawan}(1995)}]{Aryasetiawan1995b}%
  \BibitemOpen
  \bibfield  {author} {\bibinfo {author} {\bibfnamefont {K.}~\bibnamefont
  {Karlsson}}\ and\ \bibinfo {author} {\bibfnamefont {F.}~\bibnamefont
  {Aryasetiawan}},\ }\href {https://doi.org/10.1103/PhysRevB.52.4823}
  {\bibfield  {journal} {\bibinfo  {journal} {Phys. Rev. B}\ }\textbf {\bibinfo
  {volume} {52}},\ \bibinfo {pages} {4823} (\bibinfo {year}
  {1995})}\BibitemShut {NoStop}%
\bibitem [{\citenamefont {Aryasetiawan}\ \emph {et~al.}(1996)\citenamefont
  {Aryasetiawan}, \citenamefont {Hedin},\ and\ \citenamefont
  {Karlsson}}]{Aryasetiawan1996b}%
  \BibitemOpen
  \bibfield  {author} {\bibinfo {author} {\bibfnamefont {F.}~\bibnamefont
  {Aryasetiawan}}, \bibinfo {author} {\bibfnamefont {L.}~\bibnamefont
  {Hedin}},\ and\ \bibinfo {author} {\bibfnamefont {K.}~\bibnamefont
  {Karlsson}},\ }\href {https://doi.org/10.1103/PhysRevLett.77.2268} {\bibfield
   {journal} {\bibinfo  {journal} {Phys. Rev. Lett.}\ }\textbf {\bibinfo
  {volume} {77}},\ \bibinfo {pages} {2268} (\bibinfo {year}
  {1996})}\BibitemShut {NoStop}%
\bibitem [{\citenamefont {Steiner}\ \emph {et~al.}(1978)\citenamefont
  {Steiner}, \citenamefont {H{\"o}chst},\ and\ \citenamefont
  {H{\"u}fner}}]{Steiner1978}%
  \BibitemOpen
  \bibfield  {author} {\bibinfo {author} {\bibfnamefont {P.}~\bibnamefont
  {Steiner}}, \bibinfo {author} {\bibfnamefont {H.}~\bibnamefont
  {H{\"o}chst}},\ and\ \bibinfo {author} {\bibfnamefont {S.}~\bibnamefont
  {H{\"u}fner}},\ }\href {https://doi.org/10.1007/BF01320978} {\bibfield
  {journal} {\bibinfo  {journal} {Zeitschrift f{\"u}r Physik B Condensed
  Matter}\ }\textbf {\bibinfo {volume} {30}},\ \bibinfo {pages} {129} (\bibinfo
  {year} {1978})}\BibitemShut {NoStop}%
\bibitem [{\citenamefont {Campagna}\ \emph {et~al.}(1975)\citenamefont
  {Campagna}, \citenamefont {Wertheim}, \citenamefont {Shanks}, \citenamefont
  {Zumsteg},\ and\ \citenamefont {Banks}}]{Campagna1975}%
  \BibitemOpen
  \bibfield  {author} {\bibinfo {author} {\bibfnamefont {M.}~\bibnamefont
  {Campagna}}, \bibinfo {author} {\bibfnamefont {G.~K.}\ \bibnamefont
  {Wertheim}}, \bibinfo {author} {\bibfnamefont {H.~R.}\ \bibnamefont
  {Shanks}}, \bibinfo {author} {\bibfnamefont {F.}~\bibnamefont {Zumsteg}},\
  and\ \bibinfo {author} {\bibfnamefont {E.}~\bibnamefont {Banks}},\ }\href
  {https://doi.org/10.1103/PhysRevLett.34.738} {\bibfield  {journal} {\bibinfo
  {journal} {Phys. Rev. Lett.}\ }\textbf {\bibinfo {volume} {34}},\ \bibinfo
  {pages} {738} (\bibinfo {year} {1975})}\BibitemShut {NoStop}%
\bibitem [{\citenamefont {Chazalviel}\ \emph {et~al.}(1977)\citenamefont
  {Chazalviel}, \citenamefont {Campagna}, \citenamefont {Wertheim},\ and\
  \citenamefont {Shanks}}]{Chazalviel1977}%
  \BibitemOpen
  \bibfield  {author} {\bibinfo {author} {\bibfnamefont {J.~N.}\ \bibnamefont
  {Chazalviel}}, \bibinfo {author} {\bibfnamefont {M.}~\bibnamefont
  {Campagna}}, \bibinfo {author} {\bibfnamefont {G.~K.}\ \bibnamefont
  {Wertheim}},\ and\ \bibinfo {author} {\bibfnamefont {H.~R.}\ \bibnamefont
  {Shanks}},\ }\href {https://doi.org/10.1103/PhysRevB.16.697} {\bibfield
  {journal} {\bibinfo  {journal} {Phys. Rev. B}\ }\textbf {\bibinfo {volume}
  {16}},\ \bibinfo {pages} {697} (\bibinfo {year} {1977})}\BibitemShut
  {NoStop}%
\bibitem [{\citenamefont {Beatham}\ \emph {et~al.}(1980)\citenamefont
  {Beatham}, \citenamefont {Cox}, \citenamefont {Egdell},\ and\ \citenamefont
  {Orchard}}]{Beatham1980}%
  \BibitemOpen
  \bibfield  {author} {\bibinfo {author} {\bibfnamefont {N.}~\bibnamefont
  {Beatham}}, \bibinfo {author} {\bibfnamefont {P.}~\bibnamefont {Cox}},
  \bibinfo {author} {\bibfnamefont {R.}~\bibnamefont {Egdell}},\ and\ \bibinfo
  {author} {\bibfnamefont {A.}~\bibnamefont {Orchard}},\ }\href
  {https://doi.org/https://doi.org/10.1016/0009-2614(80)85108-6} {\bibfield
  {journal} {\bibinfo  {journal} {Chemical Physics Letters}\ }\textbf {\bibinfo
  {volume} {69}},\ \bibinfo {pages} {479 } (\bibinfo {year}
  {1980})}\BibitemShut {NoStop}%
\bibitem [{\citenamefont {Aryasetiawan}\ and\ \citenamefont
  {Gunnarsson}(1995)}]{Aryasetiawan1995}%
  \BibitemOpen
  \bibfield  {author} {\bibinfo {author} {\bibfnamefont {F.}~\bibnamefont
  {Aryasetiawan}}\ and\ \bibinfo {author} {\bibfnamefont {O.}~\bibnamefont
  {Gunnarsson}},\ }\href {https://doi.org/10.1103/PhysRevLett.74.3221}
  {\bibfield  {journal} {\bibinfo  {journal} {Phys. Rev. Lett.}\ }\textbf
  {\bibinfo {volume} {74}},\ \bibinfo {pages} {3221} (\bibinfo {year}
  {1995})}\BibitemShut {NoStop}%
\bibitem [{\citenamefont {Egdell}\ \emph {et~al.}(1999)\citenamefont {Egdell},
  \citenamefont {Rebane}, \citenamefont {Walker},\ and\ \citenamefont
  {Law}}]{Egdell1999}%
  \BibitemOpen
  \bibfield  {author} {\bibinfo {author} {\bibfnamefont {R.~G.}\ \bibnamefont
  {Egdell}}, \bibinfo {author} {\bibfnamefont {J.}~\bibnamefont {Rebane}},
  \bibinfo {author} {\bibfnamefont {T.~J.}\ \bibnamefont {Walker}},\ and\
  \bibinfo {author} {\bibfnamefont {D.~S.~L.}\ \bibnamefont {Law}},\ }\href
  {https://doi.org/10.1103/PhysRevB.59.1792} {\bibfield  {journal} {\bibinfo
  {journal} {Phys. Rev. B}\ }\textbf {\bibinfo {volume} {59}},\ \bibinfo
  {pages} {1792} (\bibinfo {year} {1999})}\BibitemShut {NoStop}%
\bibitem [{\citenamefont {Christou}\ \emph {et~al.}(2000)\citenamefont
  {Christou}, \citenamefont {Etchells}, \citenamefont {Renault}, \citenamefont
  {Dobson}, \citenamefont {Salata}, \citenamefont {Beamson},\ and\
  \citenamefont {Egdell}}]{Christou2000}%
  \BibitemOpen
  \bibfield  {author} {\bibinfo {author} {\bibfnamefont {V.}~\bibnamefont
  {Christou}}, \bibinfo {author} {\bibfnamefont {M.}~\bibnamefont {Etchells}},
  \bibinfo {author} {\bibfnamefont {O.}~\bibnamefont {Renault}}, \bibinfo
  {author} {\bibfnamefont {P.~J.}\ \bibnamefont {Dobson}}, \bibinfo {author}
  {\bibfnamefont {O.~V.}\ \bibnamefont {Salata}}, \bibinfo {author}
  {\bibfnamefont {G.}~\bibnamefont {Beamson}},\ and\ \bibinfo {author}
  {\bibfnamefont {R.~G.}\ \bibnamefont {Egdell}},\ }\href
  {https://doi.org/10.1063/1.1312847} {\bibfield  {journal} {\bibinfo
  {journal} {Journal of Applied Physics}\ }\textbf {\bibinfo {volume} {88}},\
  \bibinfo {pages} {5180} (\bibinfo {year} {2000})},\ \Eprint
  {https://arxiv.org/abs/https://doi.org/10.1063/1.1312847}
  {https://doi.org/10.1063/1.1312847} \BibitemShut {NoStop}%
\bibitem [{\citenamefont {Kohiki}\ \emph {et~al.}(2000)\citenamefont {Kohiki},
  \citenamefont {Arai}, \citenamefont {Yoshikawa}, \citenamefont {Fukushima},
  \citenamefont {Oku},\ and\ \citenamefont {Waseda}}]{Kohiki2000}%
  \BibitemOpen
  \bibfield  {author} {\bibinfo {author} {\bibfnamefont {S.}~\bibnamefont
  {Kohiki}}, \bibinfo {author} {\bibfnamefont {M.}~\bibnamefont {Arai}},
  \bibinfo {author} {\bibfnamefont {H.}~\bibnamefont {Yoshikawa}}, \bibinfo
  {author} {\bibfnamefont {S.}~\bibnamefont {Fukushima}}, \bibinfo {author}
  {\bibfnamefont {M.}~\bibnamefont {Oku}},\ and\ \bibinfo {author}
  {\bibfnamefont {Y.}~\bibnamefont {Waseda}},\ }\href
  {https://doi.org/10.1103/PhysRevB.62.7964} {\bibfield  {journal} {\bibinfo
  {journal} {Phys. Rev. B}\ }\textbf {\bibinfo {volume} {62}},\ \bibinfo
  {pages} {7964} (\bibinfo {year} {2000})}\BibitemShut {NoStop}%
\bibitem [{\citenamefont {Gatti}\ \emph {et~al.}(2007)\citenamefont {Gatti},
  \citenamefont {Bruneval}, \citenamefont {Olevano},\ and\ \citenamefont
  {Reining}}]{Gatti2007}%
  \BibitemOpen
  \bibfield  {author} {\bibinfo {author} {\bibfnamefont {M.}~\bibnamefont
  {Gatti}}, \bibinfo {author} {\bibfnamefont {F.}~\bibnamefont {Bruneval}},
  \bibinfo {author} {\bibfnamefont {V.}~\bibnamefont {Olevano}},\ and\ \bibinfo
  {author} {\bibfnamefont {L.}~\bibnamefont {Reining}},\ }\href
  {https://doi.org/10.1103/PhysRevLett.99.266402} {\bibfield  {journal}
  {\bibinfo  {journal} {Phys. Rev. Lett.}\ }\textbf {\bibinfo {volume} {99}},\
  \bibinfo {pages} {266402} (\bibinfo {year} {2007})}\BibitemShut {NoStop}%
\bibitem [{\citenamefont {Mudd}\ \emph {et~al.}(2014)\citenamefont {Mudd},
  \citenamefont {Lee}, \citenamefont {Mu\~noz Sanjos\'e}, \citenamefont
  {Z\'u\~niga P\'erez}, \citenamefont {Hesp}, \citenamefont {Kahk},
  \citenamefont {Payne}, \citenamefont {Egdell},\ and\ \citenamefont
  {McConville}}]{Mudd2014}%
  \BibitemOpen
  \bibfield  {author} {\bibinfo {author} {\bibfnamefont {J.~J.}\ \bibnamefont
  {Mudd}}, \bibinfo {author} {\bibfnamefont {T.-L.}\ \bibnamefont {Lee}},
  \bibinfo {author} {\bibfnamefont {V.}~\bibnamefont {Mu\~noz Sanjos\'e}},
  \bibinfo {author} {\bibfnamefont {J.}~\bibnamefont {Z\'u\~niga P\'erez}},
  \bibinfo {author} {\bibfnamefont {D.}~\bibnamefont {Hesp}}, \bibinfo {author}
  {\bibfnamefont {J.~M.}\ \bibnamefont {Kahk}}, \bibinfo {author}
  {\bibfnamefont {D.~J.}\ \bibnamefont {Payne}}, \bibinfo {author}
  {\bibfnamefont {R.~G.}\ \bibnamefont {Egdell}},\ and\ \bibinfo {author}
  {\bibfnamefont {C.~F.}\ \bibnamefont {McConville}},\ }\href
  {https://doi.org/10.1103/PhysRevB.89.035203} {\bibfield  {journal} {\bibinfo
  {journal} {Phys. Rev. B}\ }\textbf {\bibinfo {volume} {89}},\ \bibinfo
  {pages} {035203} (\bibinfo {year} {2014})}\BibitemShut {NoStop}%
\bibitem [{\citenamefont {Cox}\ \emph {et~al.}(1986)\citenamefont {Cox},
  \citenamefont {Goodenough}, \citenamefont {Tavener}, \citenamefont {Telles},\
  and\ \citenamefont {Egdell}}]{Cox1986}%
  \BibitemOpen
  \bibfield  {author} {\bibinfo {author} {\bibfnamefont {P.}~\bibnamefont
  {Cox}}, \bibinfo {author} {\bibfnamefont {J.}~\bibnamefont {Goodenough}},
  \bibinfo {author} {\bibfnamefont {P.}~\bibnamefont {Tavener}}, \bibinfo
  {author} {\bibfnamefont {D.}~\bibnamefont {Telles}},\ and\ \bibinfo {author}
  {\bibfnamefont {R.}~\bibnamefont {Egdell}},\ }\href
  {https://doi.org/https://doi.org/10.1016/0022-4596(86)90251-3} {\bibfield
  {journal} {\bibinfo  {journal} {Journal of Solid State Chemistry}\ }\textbf
  {\bibinfo {volume} {62}},\ \bibinfo {pages} {360 } (\bibinfo {year}
  {1986})}\BibitemShut {NoStop}%
\bibitem [{\citenamefont {Bozovic}(1990)}]{Bozovic1990}%
  \BibitemOpen
  \bibfield  {author} {\bibinfo {author} {\bibfnamefont {I.}~\bibnamefont
  {Bozovic}},\ }\href {https://doi.org/10.1103/PhysRevB.42.1969} {\bibfield
  {journal} {\bibinfo  {journal} {Phys. Rev. B}\ }\textbf {\bibinfo {volume}
  {42}},\ \bibinfo {pages} {1969} (\bibinfo {year} {1990})}\BibitemShut
  {NoStop}%
\bibitem [{\citenamefont {van~der Marel}(2004)}]{Marel2004}%
  \BibitemOpen
  \bibfield  {author} {\bibinfo {author} {\bibfnamefont {D.}~\bibnamefont
  {van~der Marel}},\ }\href@noop {} {\bibfield  {journal} {\bibinfo  {journal}
  {Journal of Superconductivity}\ }\textbf {\bibinfo {volume} {17}},\ \bibinfo
  {pages} {559} (\bibinfo {year} {2004})}\BibitemShut {NoStop}%
\bibitem [{\citenamefont {Werner}\ \emph {et~al.}(2015)\citenamefont {Werner},
  \citenamefont {Sakuma}, \citenamefont {Nilsson},\ and\ \citenamefont
  {Aryasetiawan}}]{Werner2015screen}%
  \BibitemOpen
  \bibfield  {author} {\bibinfo {author} {\bibfnamefont {P.}~\bibnamefont
  {Werner}}, \bibinfo {author} {\bibfnamefont {R.}~\bibnamefont {Sakuma}},
  \bibinfo {author} {\bibfnamefont {F.}~\bibnamefont {Nilsson}},\ and\ \bibinfo
  {author} {\bibfnamefont {F.}~\bibnamefont {Aryasetiawan}},\ }\href
  {https://doi.org/10.1103/PhysRevB.91.125142} {\bibfield  {journal} {\bibinfo
  {journal} {Phys. Rev. B}\ }\textbf {\bibinfo {volume} {91}},\ \bibinfo
  {pages} {125142} (\bibinfo {year} {2015})}\BibitemShut {NoStop}%
\bibitem [{\citenamefont {Luo}\ \emph {et~al.}(2013)\citenamefont {Luo},
  \citenamefont {Qiu}, \citenamefont {Lu},\ and\ \citenamefont
  {Ni}}]{Xiaoguang2013}%
  \BibitemOpen
  \bibfield  {author} {\bibinfo {author} {\bibfnamefont {X.}~\bibnamefont
  {Luo}}, \bibinfo {author} {\bibfnamefont {T.}~\bibnamefont {Qiu}}, \bibinfo
  {author} {\bibfnamefont {W.}~\bibnamefont {Lu}},\ and\ \bibinfo {author}
  {\bibfnamefont {Z.}~\bibnamefont {Ni}},\ }\href
  {https://doi.org/https://doi.org/10.1016/j.mser.2013.09.001} {\bibfield
  {journal} {\bibinfo  {journal} {Materials Science and Engineering: R:
  Reports}\ }\textbf {\bibinfo {volume} {74}},\ \bibinfo {pages} {351 }
  (\bibinfo {year} {2013})}\BibitemShut {NoStop}%
\bibitem [{\citenamefont {Guzzo}\ \emph {et~al.}(2014)\citenamefont {Guzzo},
  \citenamefont {Kas}, \citenamefont {Sponza}, \citenamefont {Giorgetti},
  \citenamefont {Sottile}, \citenamefont {Pierucci}, \citenamefont {Silly},
  \citenamefont {Sirotti}, \citenamefont {Rehr},\ and\ \citenamefont
  {Reining}}]{Guzzo2014}%
  \BibitemOpen
  \bibfield  {author} {\bibinfo {author} {\bibfnamefont {M.}~\bibnamefont
  {Guzzo}}, \bibinfo {author} {\bibfnamefont {J.~J.}\ \bibnamefont {Kas}},
  \bibinfo {author} {\bibfnamefont {L.}~\bibnamefont {Sponza}}, \bibinfo
  {author} {\bibfnamefont {C.}~\bibnamefont {Giorgetti}}, \bibinfo {author}
  {\bibfnamefont {F.}~\bibnamefont {Sottile}}, \bibinfo {author} {\bibfnamefont
  {D.}~\bibnamefont {Pierucci}}, \bibinfo {author} {\bibfnamefont {M.~G.}\
  \bibnamefont {Silly}}, \bibinfo {author} {\bibfnamefont {F.}~\bibnamefont
  {Sirotti}}, \bibinfo {author} {\bibfnamefont {J.~J.}\ \bibnamefont {Rehr}},\
  and\ \bibinfo {author} {\bibfnamefont {L.}~\bibnamefont {Reining}},\ }\href
  {https://doi.org/10.1103/PhysRevB.89.085425} {\bibfield  {journal} {\bibinfo
  {journal} {Phys. Rev. B}\ }\textbf {\bibinfo {volume} {89}},\ \bibinfo
  {pages} {085425} (\bibinfo {year} {2014})}\BibitemShut {NoStop}%
\bibitem [{\citenamefont {Raveau}(2005)}]{Raveau2005}%
  \BibitemOpen
  \bibfield  {author} {\bibinfo {author} {\bibfnamefont {B.}~\bibnamefont
  {Raveau}},\ }\href
  {https://doi.org/https://doi.org/10.1016/j.jeurceramsoc.2005.03.220}
  {\bibfield  {journal} {\bibinfo  {journal} {Journal of the European Ceramic
  Society}\ }\textbf {\bibinfo {volume} {25}},\ \bibinfo {pages} {1965}
  (\bibinfo {year} {2005})},\ \bibinfo {note} {elecroceramics IX}\BibitemShut
  {NoStop}%
\bibitem [{\citenamefont {Cheng}\ \emph {et~al.}(2011)\citenamefont {Cheng},
  \citenamefont {Liang}, \citenamefont {Tao},\ and\ \citenamefont
  {Chen}}]{Cheng2011}%
  \BibitemOpen
  \bibfield  {author} {\bibinfo {author} {\bibfnamefont {F.}~\bibnamefont
  {Cheng}}, \bibinfo {author} {\bibfnamefont {J.}~\bibnamefont {Liang}},
  \bibinfo {author} {\bibfnamefont {Z.}~\bibnamefont {Tao}},\ and\ \bibinfo
  {author} {\bibfnamefont {J.}~\bibnamefont {Chen}},\ }\href
  {https://doi.org/https://doi.org/10.1002/adma.201003587} {\bibfield
  {journal} {\bibinfo  {journal} {Advanced Materials}\ }\textbf {\bibinfo
  {volume} {23}},\ \bibinfo {pages} {1695} (\bibinfo {year} {2011})},\ \Eprint
  {https://arxiv.org/abs/https://onlinelibrary.wiley.com/doi/pdf/10.1002/adma.201003587}
  {https://onlinelibrary.wiley.com/doi/pdf/10.1002/adma.201003587} \BibitemShut
  {NoStop}%
\bibitem [{\citenamefont {Nuraje}\ \emph {et~al.}(2012)\citenamefont {Nuraje},
  \citenamefont {Asmatulu},\ and\ \citenamefont {Kudaibergenov}}]{Nuraje2012}%
  \BibitemOpen
  \bibfield  {author} {\bibinfo {author} {\bibfnamefont {N.}~\bibnamefont
  {Nuraje}}, \bibinfo {author} {\bibfnamefont {R.}~\bibnamefont {Asmatulu}},\
  and\ \bibinfo {author} {\bibfnamefont {S.}~\bibnamefont {Kudaibergenov}},\
  }\href
  {https://www.ingentaconnect.com/content/ben/cic/2012/00000002/00000002/art00004}
  {\bibfield  {journal} {\bibinfo  {journal} {Current Inorganic Chemistry}\
  }\textbf {\bibinfo {volume} {2}},\ \bibinfo {pages} {124} (\bibinfo {year}
  {2012})}\BibitemShut {NoStop}%
\bibitem [{\citenamefont {Hou}\ and\ \citenamefont {Cronin}(2013)}]{Hou2013}%
  \BibitemOpen
  \bibfield  {author} {\bibinfo {author} {\bibfnamefont {W.}~\bibnamefont
  {Hou}}\ and\ \bibinfo {author} {\bibfnamefont {S.~B.}\ \bibnamefont
  {Cronin}},\ }\href {https://doi.org/https://doi.org/10.1002/adfm.201202148}
  {\bibfield  {journal} {\bibinfo  {journal} {Advanced Functional Materials}\
  }\textbf {\bibinfo {volume} {23}},\ \bibinfo {pages} {1612} (\bibinfo {year}
  {2013})},\ \Eprint
  {https://arxiv.org/abs/https://onlinelibrary.wiley.com/doi/pdf/10.1002/adfm.201202148}
  {https://onlinelibrary.wiley.com/doi/pdf/10.1002/adfm.201202148} \BibitemShut
  {NoStop}%
\bibitem [{\citenamefont {Meng}\ \emph {et~al.}(2016)\citenamefont {Meng},
  \citenamefont {Liu}, \citenamefont {Ouyang}, \citenamefont {Xu},
  \citenamefont {Wang}, \citenamefont {Zhao},\ and\ \citenamefont
  {Ye}}]{Meng2016}%
  \BibitemOpen
  \bibfield  {author} {\bibinfo {author} {\bibfnamefont {X.}~\bibnamefont
  {Meng}}, \bibinfo {author} {\bibfnamefont {L.}~\bibnamefont {Liu}}, \bibinfo
  {author} {\bibfnamefont {S.}~\bibnamefont {Ouyang}}, \bibinfo {author}
  {\bibfnamefont {H.}~\bibnamefont {Xu}}, \bibinfo {author} {\bibfnamefont
  {D.}~\bibnamefont {Wang}}, \bibinfo {author} {\bibfnamefont {N.}~\bibnamefont
  {Zhao}},\ and\ \bibinfo {author} {\bibfnamefont {J.}~\bibnamefont {Ye}},\
  }\href {https://doi.org/https://doi.org/10.1002/adma.201600305} {\bibfield
  {journal} {\bibinfo  {journal} {Advanced Materials}\ }\textbf {\bibinfo
  {volume} {28}},\ \bibinfo {pages} {6781} (\bibinfo {year} {2016})},\ \Eprint
  {https://arxiv.org/abs/https://onlinelibrary.wiley.com/doi/pdf/10.1002/adma.201600305}
  {https://onlinelibrary.wiley.com/doi/pdf/10.1002/adma.201600305} \BibitemShut
  {NoStop}%
\bibitem [{\citenamefont {Szunerits}\ and\ \citenamefont
  {Boukherroub}(2012)}]{Szunerits2012}%
  \BibitemOpen
  \bibfield  {author} {\bibinfo {author} {\bibfnamefont {S.}~\bibnamefont
  {Szunerits}}\ and\ \bibinfo {author} {\bibfnamefont {R.}~\bibnamefont
  {Boukherroub}},\ }\href {https://doi.org/10.1039/C2CC33266C} {\bibfield
  {journal} {\bibinfo  {journal} {Chem. Commun.}\ }\textbf {\bibinfo {volume}
  {48}},\ \bibinfo {pages} {8999} (\bibinfo {year} {2012})}\BibitemShut
  {NoStop}%
\bibitem [{\citenamefont {Wang}\ \emph {et~al.}(1996)\citenamefont {Wang},
  \citenamefont {Zhang}, \citenamefont {Dravid}, \citenamefont {Ng},
  \citenamefont {Klein}, \citenamefont {Schnatterly},\ and\ \citenamefont
  {Miller}}]{Wang1996}%
  \BibitemOpen
  \bibfield  {author} {\bibinfo {author} {\bibfnamefont {Y.~Y.}\ \bibnamefont
  {Wang}}, \bibinfo {author} {\bibfnamefont {F.~C.}\ \bibnamefont {Zhang}},
  \bibinfo {author} {\bibfnamefont {V.~P.}\ \bibnamefont {Dravid}}, \bibinfo
  {author} {\bibfnamefont {K.~K.}\ \bibnamefont {Ng}}, \bibinfo {author}
  {\bibfnamefont {M.~V.}\ \bibnamefont {Klein}}, \bibinfo {author}
  {\bibfnamefont {S.~E.}\ \bibnamefont {Schnatterly}},\ and\ \bibinfo {author}
  {\bibfnamefont {L.~L.}\ \bibnamefont {Miller}},\ }\href
  {https://doi.org/10.1103/PhysRevLett.77.1809} {\bibfield  {journal} {\bibinfo
   {journal} {Phys. Rev. Lett.}\ }\textbf {\bibinfo {volume} {77}},\ \bibinfo
  {pages} {1809} (\bibinfo {year} {1996})}\BibitemShut {NoStop}%
\bibitem [{\citenamefont {Makino}\ \emph {et~al.}(1998)\citenamefont {Makino},
  \citenamefont {Inoue}, \citenamefont {Rozenberg}, \citenamefont {Hase},
  \citenamefont {Aiura},\ and\ \citenamefont {Onari}}]{Onari1998}%
  \BibitemOpen
  \bibfield  {author} {\bibinfo {author} {\bibfnamefont {H.}~\bibnamefont
  {Makino}}, \bibinfo {author} {\bibfnamefont {I.~H.}\ \bibnamefont {Inoue}},
  \bibinfo {author} {\bibfnamefont {M.~J.}\ \bibnamefont {Rozenberg}}, \bibinfo
  {author} {\bibfnamefont {I.}~\bibnamefont {Hase}}, \bibinfo {author}
  {\bibfnamefont {Y.}~\bibnamefont {Aiura}},\ and\ \bibinfo {author}
  {\bibfnamefont {S.}~\bibnamefont {Onari}},\ }\href
  {https://doi.org/10.1103/PhysRevB.58.4384} {\bibfield  {journal} {\bibinfo
  {journal} {Phys. Rev. B}\ }\textbf {\bibinfo {volume} {58}},\ \bibinfo
  {pages} {4384} (\bibinfo {year} {1998})}\BibitemShut {NoStop}%
\bibitem [{\citenamefont {Grosvenor}\ \emph {et~al.}(2006)\citenamefont
  {Grosvenor}, \citenamefont {Biesinger}, \citenamefont {Smart},\ and\
  \citenamefont {McIntyre}}]{Grosvenor2006}%
  \BibitemOpen
  \bibfield  {author} {\bibinfo {author} {\bibfnamefont {A.~P.}\ \bibnamefont
  {Grosvenor}}, \bibinfo {author} {\bibfnamefont {M.~C.}\ \bibnamefont
  {Biesinger}}, \bibinfo {author} {\bibfnamefont {R.~S.}\ \bibnamefont
  {Smart}},\ and\ \bibinfo {author} {\bibfnamefont {N.~S.}\ \bibnamefont
  {McIntyre}},\ }\href
  {https://doi.org/https://doi.org/10.1016/j.susc.2006.01.041} {\bibfield
  {journal} {\bibinfo  {journal} {Surface Science}\ }\textbf {\bibinfo {volume}
  {600}},\ \bibinfo {pages} {1771 } (\bibinfo {year} {2006})}\BibitemShut
  {NoStop}%
\bibitem [{\citenamefont {Markiewicz}\ and\ \citenamefont
  {Bansil}(2007)}]{Bansil2007}%
  \BibitemOpen
  \bibfield  {author} {\bibinfo {author} {\bibfnamefont {R.~S.}\ \bibnamefont
  {Markiewicz}}\ and\ \bibinfo {author} {\bibfnamefont {A.}~\bibnamefont
  {Bansil}},\ }\href {https://doi.org/10.1103/PhysRevB.75.020508} {\bibfield
  {journal} {\bibinfo  {journal} {Phys. Rev. B}\ }\textbf {\bibinfo {volume}
  {75}},\ \bibinfo {pages} {020508} (\bibinfo {year} {2007})}\BibitemShut
  {NoStop}%
\bibitem [{\citenamefont {Gatti}\ and\ \citenamefont
  {Guzzo}(2013)}]{Gatti2013}%
  \BibitemOpen
  \bibfield  {author} {\bibinfo {author} {\bibfnamefont {M.}~\bibnamefont
  {Gatti}}\ and\ \bibinfo {author} {\bibfnamefont {M.}~\bibnamefont {Guzzo}},\
  }\href {https://doi.org/10.1103/PhysRevB.87.155147} {\bibfield  {journal}
  {\bibinfo  {journal} {Phys. Rev. B}\ }\textbf {\bibinfo {volume} {87}},\
  \bibinfo {pages} {155147} (\bibinfo {year} {2013})}\BibitemShut {NoStop}%
\bibitem [{\citenamefont {Boehnke}\ \emph {et~al.}(2016)\citenamefont
  {Boehnke}, \citenamefont {Nilsson}, \citenamefont {Aryasetiawan},\ and\
  \citenamefont {Werner}}]{Boehnke2016}%
  \BibitemOpen
  \bibfield  {author} {\bibinfo {author} {\bibfnamefont {L.}~\bibnamefont
  {Boehnke}}, \bibinfo {author} {\bibfnamefont {F.}~\bibnamefont {Nilsson}},
  \bibinfo {author} {\bibfnamefont {F.}~\bibnamefont {Aryasetiawan}},\ and\
  \bibinfo {author} {\bibfnamefont {P.}~\bibnamefont {Werner}},\ }\href
  {https://doi.org/10.1103/PhysRevB.94.201106} {\bibfield  {journal} {\bibinfo
  {journal} {Phys. Rev. B}\ }\textbf {\bibinfo {volume} {94}},\ \bibinfo
  {pages} {201106(R)} (\bibinfo {year} {2016})}\BibitemShut {NoStop}%
\bibitem [{\citenamefont {Nilsson}\ \emph {et~al.}(2017)\citenamefont
  {Nilsson}, \citenamefont {Boehnke}, \citenamefont {Werner},\ and\
  \citenamefont {Aryasetiawan}}]{Nilsson2017}%
  \BibitemOpen
  \bibfield  {author} {\bibinfo {author} {\bibfnamefont {F.}~\bibnamefont
  {Nilsson}}, \bibinfo {author} {\bibfnamefont {L.}~\bibnamefont {Boehnke}},
  \bibinfo {author} {\bibfnamefont {P.}~\bibnamefont {Werner}},\ and\ \bibinfo
  {author} {\bibfnamefont {F.}~\bibnamefont {Aryasetiawan}},\ }\href
  {https://doi.org/10.1103/PhysRevMaterials.1.043803} {\bibfield  {journal}
  {\bibinfo  {journal} {Phys. Rev. Materials}\ }\textbf {\bibinfo {volume}
  {1}},\ \bibinfo {pages} {043803} (\bibinfo {year} {2017})}\BibitemShut
  {NoStop}%
\bibitem [{\citenamefont {Petocchi}\ \emph {et~al.}(2020)\citenamefont
  {Petocchi}, \citenamefont {Nilsson}, \citenamefont {Aryasetiawan},\ and\
  \citenamefont {Werner}}]{Petocchi2020}%
  \BibitemOpen
  \bibfield  {author} {\bibinfo {author} {\bibfnamefont {F.}~\bibnamefont
  {Petocchi}}, \bibinfo {author} {\bibfnamefont {F.}~\bibnamefont {Nilsson}},
  \bibinfo {author} {\bibfnamefont {F.}~\bibnamefont {Aryasetiawan}},\ and\
  \bibinfo {author} {\bibfnamefont {P.}~\bibnamefont {Werner}},\ }\href
  {https://doi.org/10.1103/PhysRevResearch.2.013191} {\bibfield  {journal}
  {\bibinfo  {journal} {Phys. Rev. Research}\ }\textbf {\bibinfo {volume}
  {2}},\ \bibinfo {pages} {013191} (\bibinfo {year} {2020})}\BibitemShut
  {NoStop}%
\bibitem [{\citenamefont {Yeh}\ \emph {et~al.}(2021)\citenamefont {Yeh},
  \citenamefont {Iskakov}, \citenamefont {Zgid},\ and\ \citenamefont
  {Gull}}]{Yeh2021}%
  \BibitemOpen
  \bibfield  {author} {\bibinfo {author} {\bibfnamefont {C.-N.}\ \bibnamefont
  {Yeh}}, \bibinfo {author} {\bibfnamefont {S.}~\bibnamefont {Iskakov}},
  \bibinfo {author} {\bibfnamefont {D.}~\bibnamefont {Zgid}},\ and\ \bibinfo
  {author} {\bibfnamefont {E.}~\bibnamefont {Gull}},\ }\href
  {https://doi.org/10.1103/PhysRevB.103.195149} {\bibfield  {journal} {\bibinfo
   {journal} {Phys. Rev. B}\ }\textbf {\bibinfo {volume} {103}},\ \bibinfo
  {pages} {195149} (\bibinfo {year} {2021})}\BibitemShut {NoStop}%
\bibitem [{\citenamefont {Inoue}\ \emph {et~al.}(1995)\citenamefont {Inoue},
  \citenamefont {Hase}, \citenamefont {Aiura}, \citenamefont {Fujimori},
  \citenamefont {Haruyama}, \citenamefont {Maruyama},\ and\ \citenamefont
  {Nishihara}}]{Inoue1995}%
  \BibitemOpen
  \bibfield  {author} {\bibinfo {author} {\bibfnamefont {I.~H.}\ \bibnamefont
  {Inoue}}, \bibinfo {author} {\bibfnamefont {I.}~\bibnamefont {Hase}},
  \bibinfo {author} {\bibfnamefont {Y.}~\bibnamefont {Aiura}}, \bibinfo
  {author} {\bibfnamefont {A.}~\bibnamefont {Fujimori}}, \bibinfo {author}
  {\bibfnamefont {Y.}~\bibnamefont {Haruyama}}, \bibinfo {author}
  {\bibfnamefont {T.}~\bibnamefont {Maruyama}},\ and\ \bibinfo {author}
  {\bibfnamefont {Y.}~\bibnamefont {Nishihara}},\ }\href
  {https://doi.org/10.1103/PhysRevLett.74.2539} {\bibfield  {journal} {\bibinfo
   {journal} {Phys. Rev. Lett.}\ }\textbf {\bibinfo {volume} {74}},\ \bibinfo
  {pages} {2539} (\bibinfo {year} {1995})}\BibitemShut {NoStop}%
\bibitem [{\citenamefont {Rozenberg}\ \emph {et~al.}(1996)\citenamefont
  {Rozenberg}, \citenamefont {Inoue}, \citenamefont {Makino}, \citenamefont
  {Iga},\ and\ \citenamefont {Nishihara}}]{Rozenberg1996}%
  \BibitemOpen
  \bibfield  {author} {\bibinfo {author} {\bibfnamefont {M.~J.}\ \bibnamefont
  {Rozenberg}}, \bibinfo {author} {\bibfnamefont {I.~H.}\ \bibnamefont
  {Inoue}}, \bibinfo {author} {\bibfnamefont {H.}~\bibnamefont {Makino}},
  \bibinfo {author} {\bibfnamefont {F.}~\bibnamefont {Iga}},\ and\ \bibinfo
  {author} {\bibfnamefont {Y.}~\bibnamefont {Nishihara}},\ }\href
  {https://doi.org/10.1103/PhysRevLett.76.4781} {\bibfield  {journal} {\bibinfo
   {journal} {Phys. Rev. Lett.}\ }\textbf {\bibinfo {volume} {76}},\ \bibinfo
  {pages} {4781} (\bibinfo {year} {1996})}\BibitemShut {NoStop}%
\bibitem [{\citenamefont {Sekiyama}\ \emph {et~al.}(2004)\citenamefont
  {Sekiyama}, \citenamefont {Fujiwara}, \citenamefont {Imada}, \citenamefont
  {Suga}, \citenamefont {Eisaki}, \citenamefont {Uchida}, \citenamefont
  {Takegahara}, \citenamefont {Harima}, \citenamefont {Saitoh}, \citenamefont
  {Nekrasov}, \citenamefont {Keller}, \citenamefont {Kondakov}, \citenamefont
  {Kozhevnikov}, \citenamefont {Pruschke}, \citenamefont {Held}, \citenamefont
  {Vollhardt},\ and\ \citenamefont {Anisimov}}]{Sekiyama2004}%
  \BibitemOpen
  \bibfield  {author} {\bibinfo {author} {\bibfnamefont {A.}~\bibnamefont
  {Sekiyama}}, \bibinfo {author} {\bibfnamefont {H.}~\bibnamefont {Fujiwara}},
  \bibinfo {author} {\bibfnamefont {S.}~\bibnamefont {Imada}}, \bibinfo
  {author} {\bibfnamefont {S.}~\bibnamefont {Suga}}, \bibinfo {author}
  {\bibfnamefont {H.}~\bibnamefont {Eisaki}}, \bibinfo {author} {\bibfnamefont
  {S.~I.}\ \bibnamefont {Uchida}}, \bibinfo {author} {\bibfnamefont
  {K.}~\bibnamefont {Takegahara}}, \bibinfo {author} {\bibfnamefont
  {H.}~\bibnamefont {Harima}}, \bibinfo {author} {\bibfnamefont
  {Y.}~\bibnamefont {Saitoh}}, \bibinfo {author} {\bibfnamefont {I.~A.}\
  \bibnamefont {Nekrasov}}, \bibinfo {author} {\bibfnamefont {G.}~\bibnamefont
  {Keller}}, \bibinfo {author} {\bibfnamefont {D.~E.}\ \bibnamefont
  {Kondakov}}, \bibinfo {author} {\bibfnamefont {A.~V.}\ \bibnamefont
  {Kozhevnikov}}, \bibinfo {author} {\bibfnamefont {T.}~\bibnamefont
  {Pruschke}}, \bibinfo {author} {\bibfnamefont {K.}~\bibnamefont {Held}},
  \bibinfo {author} {\bibfnamefont {D.}~\bibnamefont {Vollhardt}},\ and\
  \bibinfo {author} {\bibfnamefont {V.~I.}\ \bibnamefont {Anisimov}},\ }\href
  {https://doi.org/10.1103/PhysRevLett.93.156402} {\bibfield  {journal}
  {\bibinfo  {journal} {Phys. Rev. Lett.}\ }\textbf {\bibinfo {volume} {93}},\
  \bibinfo {pages} {156402} (\bibinfo {year} {2004})}\BibitemShut {NoStop}%
\bibitem [{\citenamefont {Takizawa}\ \emph {et~al.}(2009)\citenamefont
  {Takizawa}, \citenamefont {Minohara}, \citenamefont {Kumigashira},
  \citenamefont {Toyota}, \citenamefont {Oshima}, \citenamefont {Wadati},
  \citenamefont {Yoshida}, \citenamefont {Fujimori}, \citenamefont {Lippmaa},
  \citenamefont {Kawasaki}, \citenamefont {Koinuma}, \citenamefont {Sordi},\
  and\ \citenamefont {Rozenberg}}]{Takizawa2009}%
  \BibitemOpen
  \bibfield  {author} {\bibinfo {author} {\bibfnamefont {M.}~\bibnamefont
  {Takizawa}}, \bibinfo {author} {\bibfnamefont {M.}~\bibnamefont {Minohara}},
  \bibinfo {author} {\bibfnamefont {H.}~\bibnamefont {Kumigashira}}, \bibinfo
  {author} {\bibfnamefont {D.}~\bibnamefont {Toyota}}, \bibinfo {author}
  {\bibfnamefont {M.}~\bibnamefont {Oshima}}, \bibinfo {author} {\bibfnamefont
  {H.}~\bibnamefont {Wadati}}, \bibinfo {author} {\bibfnamefont
  {T.}~\bibnamefont {Yoshida}}, \bibinfo {author} {\bibfnamefont
  {A.}~\bibnamefont {Fujimori}}, \bibinfo {author} {\bibfnamefont
  {M.}~\bibnamefont {Lippmaa}}, \bibinfo {author} {\bibfnamefont
  {M.}~\bibnamefont {Kawasaki}}, \bibinfo {author} {\bibfnamefont
  {H.}~\bibnamefont {Koinuma}}, \bibinfo {author} {\bibfnamefont
  {G.}~\bibnamefont {Sordi}},\ and\ \bibinfo {author} {\bibfnamefont
  {M.}~\bibnamefont {Rozenberg}},\ }\href
  {https://doi.org/10.1103/PhysRevB.80.235104} {\bibfield  {journal} {\bibinfo
  {journal} {Phys. Rev. B}\ }\textbf {\bibinfo {volume} {80}},\ \bibinfo
  {pages} {235104} (\bibinfo {year} {2009})}\BibitemShut {NoStop}%
\bibitem [{\citenamefont {Aizaki}\ \emph {et~al.}(2012)\citenamefont {Aizaki},
  \citenamefont {Yoshida}, \citenamefont {Yoshimatsu}, \citenamefont
  {Takizawa}, \citenamefont {Minohara}, \citenamefont {Ideta}, \citenamefont
  {Fujimori}, \citenamefont {Gupta}, \citenamefont {Mahadevan}, \citenamefont
  {Horiba}, \citenamefont {Kumigashira},\ and\ \citenamefont
  {Oshima}}]{Aizaki2012}%
  \BibitemOpen
  \bibfield  {author} {\bibinfo {author} {\bibfnamefont {S.}~\bibnamefont
  {Aizaki}}, \bibinfo {author} {\bibfnamefont {T.}~\bibnamefont {Yoshida}},
  \bibinfo {author} {\bibfnamefont {K.}~\bibnamefont {Yoshimatsu}}, \bibinfo
  {author} {\bibfnamefont {M.}~\bibnamefont {Takizawa}}, \bibinfo {author}
  {\bibfnamefont {M.}~\bibnamefont {Minohara}}, \bibinfo {author}
  {\bibfnamefont {S.}~\bibnamefont {Ideta}}, \bibinfo {author} {\bibfnamefont
  {A.}~\bibnamefont {Fujimori}}, \bibinfo {author} {\bibfnamefont
  {K.}~\bibnamefont {Gupta}}, \bibinfo {author} {\bibfnamefont
  {P.}~\bibnamefont {Mahadevan}}, \bibinfo {author} {\bibfnamefont
  {K.}~\bibnamefont {Horiba}}, \bibinfo {author} {\bibfnamefont
  {H.}~\bibnamefont {Kumigashira}},\ and\ \bibinfo {author} {\bibfnamefont
  {M.}~\bibnamefont {Oshima}},\ }\href
  {https://doi.org/10.1103/PhysRevLett.109.056401} {\bibfield  {journal}
  {\bibinfo  {journal} {Phys. Rev. Lett.}\ }\textbf {\bibinfo {volume} {109}},\
  \bibinfo {pages} {056401} (\bibinfo {year} {2012})}\BibitemShut {NoStop}%
\bibitem [{\citenamefont {Backes}\ \emph {et~al.}(2016)\citenamefont {Backes},
  \citenamefont {R\"odel}, \citenamefont {Fortuna}, \citenamefont
  {Frantzeskakis}, \citenamefont {Le~F\`evre}, \citenamefont {Bertran},
  \citenamefont {Kobayashi}, \citenamefont {Yukawa}, \citenamefont
  {Mitsuhashi}, \citenamefont {Kitamura}, \citenamefont {Horiba}, \citenamefont
  {Kumigashira}, \citenamefont {Saint-Martin}, \citenamefont {Fouchet},
  \citenamefont {Berini}, \citenamefont {Dumont}, \citenamefont {Kim},
  \citenamefont {Lechermann}, \citenamefont {Jeschke}, \citenamefont
  {Rozenberg}, \citenamefont {Valent\'{\i}},\ and\ \citenamefont
  {Santander-Syro}}]{Backes2016}%
  \BibitemOpen
  \bibfield  {author} {\bibinfo {author} {\bibfnamefont {S.}~\bibnamefont
  {Backes}}, \bibinfo {author} {\bibfnamefont {T.~C.}\ \bibnamefont {R\"odel}},
  \bibinfo {author} {\bibfnamefont {F.}~\bibnamefont {Fortuna}}, \bibinfo
  {author} {\bibfnamefont {E.}~\bibnamefont {Frantzeskakis}}, \bibinfo {author}
  {\bibfnamefont {P.}~\bibnamefont {Le~F\`evre}}, \bibinfo {author}
  {\bibfnamefont {F.}~\bibnamefont {Bertran}}, \bibinfo {author} {\bibfnamefont
  {M.}~\bibnamefont {Kobayashi}}, \bibinfo {author} {\bibfnamefont
  {R.}~\bibnamefont {Yukawa}}, \bibinfo {author} {\bibfnamefont
  {T.}~\bibnamefont {Mitsuhashi}}, \bibinfo {author} {\bibfnamefont
  {M.}~\bibnamefont {Kitamura}}, \bibinfo {author} {\bibfnamefont
  {K.}~\bibnamefont {Horiba}}, \bibinfo {author} {\bibfnamefont
  {H.}~\bibnamefont {Kumigashira}}, \bibinfo {author} {\bibfnamefont
  {R.}~\bibnamefont {Saint-Martin}}, \bibinfo {author} {\bibfnamefont
  {A.}~\bibnamefont {Fouchet}}, \bibinfo {author} {\bibfnamefont
  {B.}~\bibnamefont {Berini}}, \bibinfo {author} {\bibfnamefont
  {Y.}~\bibnamefont {Dumont}}, \bibinfo {author} {\bibfnamefont {A.~J.}\
  \bibnamefont {Kim}}, \bibinfo {author} {\bibfnamefont {F.}~\bibnamefont
  {Lechermann}}, \bibinfo {author} {\bibfnamefont {H.~O.}\ \bibnamefont
  {Jeschke}}, \bibinfo {author} {\bibfnamefont {M.~J.}\ \bibnamefont
  {Rozenberg}}, \bibinfo {author} {\bibfnamefont {R.}~\bibnamefont
  {Valent\'{\i}}},\ and\ \bibinfo {author} {\bibfnamefont {A.~F.}\ \bibnamefont
  {Santander-Syro}},\ }\href {https://doi.org/10.1103/PhysRevB.94.241110}
  {\bibfield  {journal} {\bibinfo  {journal} {Phys. Rev. B}\ }\textbf {\bibinfo
  {volume} {94}},\ \bibinfo {pages} {241110} (\bibinfo {year}
  {2016})}\BibitemShut {NoStop}%
\bibitem [{\citenamefont {Pavarini}\ \emph {et~al.}(2004)\citenamefont
  {Pavarini}, \citenamefont {Biermann}, \citenamefont {Poteryaev},
  \citenamefont {Lichtenstein}, \citenamefont {Georges},\ and\ \citenamefont
  {Andersen}}]{Pavarini2004}%
  \BibitemOpen
  \bibfield  {author} {\bibinfo {author} {\bibfnamefont {E.}~\bibnamefont
  {Pavarini}}, \bibinfo {author} {\bibfnamefont {S.}~\bibnamefont {Biermann}},
  \bibinfo {author} {\bibfnamefont {A.}~\bibnamefont {Poteryaev}}, \bibinfo
  {author} {\bibfnamefont {A.~I.}\ \bibnamefont {Lichtenstein}}, \bibinfo
  {author} {\bibfnamefont {A.}~\bibnamefont {Georges}},\ and\ \bibinfo {author}
  {\bibfnamefont {O.~K.}\ \bibnamefont {Andersen}},\ }\href
  {https://doi.org/10.1103/PhysRevLett.92.176403} {\bibfield  {journal}
  {\bibinfo  {journal} {Phys. Rev. Lett.}\ }\textbf {\bibinfo {volume} {92}},\
  \bibinfo {pages} {176403} (\bibinfo {year} {2004})}\BibitemShut {NoStop}%
\bibitem [{\citenamefont {Amadon}\ \emph {et~al.}(2008)\citenamefont {Amadon},
  \citenamefont {Lechermann}, \citenamefont {Georges}, \citenamefont {Jollet},
  \citenamefont {Wehling},\ and\ \citenamefont {Lichtenstein}}]{Amadon2008}%
  \BibitemOpen
  \bibfield  {author} {\bibinfo {author} {\bibfnamefont {B.}~\bibnamefont
  {Amadon}}, \bibinfo {author} {\bibfnamefont {F.}~\bibnamefont {Lechermann}},
  \bibinfo {author} {\bibfnamefont {A.}~\bibnamefont {Georges}}, \bibinfo
  {author} {\bibfnamefont {F.}~\bibnamefont {Jollet}}, \bibinfo {author}
  {\bibfnamefont {T.~O.}\ \bibnamefont {Wehling}},\ and\ \bibinfo {author}
  {\bibfnamefont {A.~I.}\ \bibnamefont {Lichtenstein}},\ }\href
  {https://doi.org/10.1103/PhysRevB.77.205112} {\bibfield  {journal} {\bibinfo
  {journal} {Phys. Rev. B}\ }\textbf {\bibinfo {volume} {77}},\ \bibinfo
  {pages} {205112} (\bibinfo {year} {2008})}\BibitemShut {NoStop}%
\bibitem [{\citenamefont {Tomczak}\ \emph {et~al.}(2014)\citenamefont
  {Tomczak}, \citenamefont {Casula}, \citenamefont {Miyake},\ and\
  \citenamefont {Biermann}}]{Tomczak2014}%
  \BibitemOpen
  \bibfield  {author} {\bibinfo {author} {\bibfnamefont {J.~M.}\ \bibnamefont
  {Tomczak}}, \bibinfo {author} {\bibfnamefont {M.}~\bibnamefont {Casula}},
  \bibinfo {author} {\bibfnamefont {T.}~\bibnamefont {Miyake}},\ and\ \bibinfo
  {author} {\bibfnamefont {S.}~\bibnamefont {Biermann}},\ }\href
  {https://doi.org/10.1103/PhysRevB.90.165138} {\bibfield  {journal} {\bibinfo
  {journal} {Phys. Rev. B}\ }\textbf {\bibinfo {volume} {90}},\ \bibinfo
  {pages} {165138} (\bibinfo {year} {2014})}\BibitemShut {NoStop}%
\bibitem [{\citenamefont {Liebsch}(2003)}]{Liebsch2003}%
  \BibitemOpen
  \bibfield  {author} {\bibinfo {author} {\bibfnamefont {A.}~\bibnamefont
  {Liebsch}},\ }\href {https://doi.org/10.1103/PhysRevLett.90.096401}
  {\bibfield  {journal} {\bibinfo  {journal} {Phys. Rev. Lett.}\ }\textbf
  {\bibinfo {volume} {90}},\ \bibinfo {pages} {096401} (\bibinfo {year}
  {2003})}\BibitemShut {NoStop}%
\bibitem [{\citenamefont {Nekrasov}\ \emph {et~al.}(2005)\citenamefont
  {Nekrasov}, \citenamefont {Keller}, \citenamefont {Kondakov}, \citenamefont
  {Kozhevnikov}, \citenamefont {Pruschke}, \citenamefont {Held}, \citenamefont
  {Vollhardt},\ and\ \citenamefont {Anisimov}}]{Nekrasov2005}%
  \BibitemOpen
  \bibfield  {author} {\bibinfo {author} {\bibfnamefont {I.~A.}\ \bibnamefont
  {Nekrasov}}, \bibinfo {author} {\bibfnamefont {G.}~\bibnamefont {Keller}},
  \bibinfo {author} {\bibfnamefont {D.~E.}\ \bibnamefont {Kondakov}}, \bibinfo
  {author} {\bibfnamefont {A.~V.}\ \bibnamefont {Kozhevnikov}}, \bibinfo
  {author} {\bibfnamefont {T.}~\bibnamefont {Pruschke}}, \bibinfo {author}
  {\bibfnamefont {K.}~\bibnamefont {Held}}, \bibinfo {author} {\bibfnamefont
  {D.}~\bibnamefont {Vollhardt}},\ and\ \bibinfo {author} {\bibfnamefont
  {V.~I.}\ \bibnamefont {Anisimov}},\ }\href
  {https://doi.org/10.1103/PhysRevB.72.155106} {\bibfield  {journal} {\bibinfo
  {journal} {Phys. Rev. B}\ }\textbf {\bibinfo {volume} {72}},\ \bibinfo
  {pages} {155106} (\bibinfo {year} {2005})}\BibitemShut {NoStop}%
\bibitem [{\citenamefont {Nekrasov}\ \emph {et~al.}(2006)\citenamefont
  {Nekrasov}, \citenamefont {Held}, \citenamefont {Keller}, \citenamefont
  {Kondakov}, \citenamefont {Pruschke}, \citenamefont {Kollar}, \citenamefont
  {Andersen}, \citenamefont {Anisimov},\ and\ \citenamefont
  {Vollhardt}}]{Nekrasov2006}%
  \BibitemOpen
  \bibfield  {author} {\bibinfo {author} {\bibfnamefont {I.~A.}\ \bibnamefont
  {Nekrasov}}, \bibinfo {author} {\bibfnamefont {K.}~\bibnamefont {Held}},
  \bibinfo {author} {\bibfnamefont {G.}~\bibnamefont {Keller}}, \bibinfo
  {author} {\bibfnamefont {D.~E.}\ \bibnamefont {Kondakov}}, \bibinfo {author}
  {\bibfnamefont {T.}~\bibnamefont {Pruschke}}, \bibinfo {author}
  {\bibfnamefont {M.}~\bibnamefont {Kollar}}, \bibinfo {author} {\bibfnamefont
  {O.~K.}\ \bibnamefont {Andersen}}, \bibinfo {author} {\bibfnamefont {V.~I.}\
  \bibnamefont {Anisimov}},\ and\ \bibinfo {author} {\bibfnamefont
  {D.}~\bibnamefont {Vollhardt}},\ }\href
  {https://doi.org/10.1103/PhysRevB.73.155112} {\bibfield  {journal} {\bibinfo
  {journal} {Phys. Rev. B}\ }\textbf {\bibinfo {volume} {73}},\ \bibinfo
  {pages} {155112} (\bibinfo {year} {2006})}\BibitemShut {NoStop}%
\bibitem [{\citenamefont {Biermann}\ \emph {et~al.}(2003)\citenamefont
  {Biermann}, \citenamefont {Aryasetiawan},\ and\ \citenamefont
  {Georges}}]{Biermann2003}%
  \BibitemOpen
  \bibfield  {author} {\bibinfo {author} {\bibfnamefont {S.}~\bibnamefont
  {Biermann}}, \bibinfo {author} {\bibfnamefont {F.}~\bibnamefont
  {Aryasetiawan}},\ and\ \bibinfo {author} {\bibfnamefont {A.}~\bibnamefont
  {Georges}},\ }\href {https://doi.org/10.1103/PhysRevLett.90.086402}
  {\bibfield  {journal} {\bibinfo  {journal} {Phys. Rev. Lett.}\ }\textbf
  {\bibinfo {volume} {90}},\ \bibinfo {pages} {086402} (\bibinfo {year}
  {2003})}\BibitemShut {NoStop}%
\bibitem [{\citenamefont {Tomczak}\ \emph {et~al.}(2012)\citenamefont
  {Tomczak}, \citenamefont {Casula}, \citenamefont {Miyake}, \citenamefont
  {Aryasetiawan},\ and\ \citenamefont {Biermann}}]{Tomczak2012}%
  \BibitemOpen
  \bibfield  {author} {\bibinfo {author} {\bibfnamefont {J.~M.}\ \bibnamefont
  {Tomczak}}, \bibinfo {author} {\bibfnamefont {M.}~\bibnamefont {Casula}},
  \bibinfo {author} {\bibfnamefont {T.}~\bibnamefont {Miyake}}, \bibinfo
  {author} {\bibfnamefont {F.}~\bibnamefont {Aryasetiawan}},\ and\ \bibinfo
  {author} {\bibfnamefont {S.}~\bibnamefont {Biermann}},\ }\href
  {http://stacks.iop.org/0295-5075/100/i=6/a=67001} {\bibfield  {journal}
  {\bibinfo  {journal} {EPL (Europhysics Letters)}\ }\textbf {\bibinfo {volume}
  {100}},\ \bibinfo {pages} {67001} (\bibinfo {year} {2012})}\BibitemShut
  {NoStop}%
\bibitem [{\citenamefont {Aryasetiawan}\ \emph {et~al.}(2006)\citenamefont
  {Aryasetiawan}, \citenamefont {Karlsson}, \citenamefont {Jepsen},\ and\
  \citenamefont {Sch\"onberger}}]{Aryasetiawan2006}%
  \BibitemOpen
  \bibfield  {author} {\bibinfo {author} {\bibfnamefont {F.}~\bibnamefont
  {Aryasetiawan}}, \bibinfo {author} {\bibfnamefont {K.}~\bibnamefont
  {Karlsson}}, \bibinfo {author} {\bibfnamefont {O.}~\bibnamefont {Jepsen}},\
  and\ \bibinfo {author} {\bibfnamefont {U.}~\bibnamefont {Sch\"onberger}},\
  }\href {https://doi.org/10.1103/PhysRevB.74.125106} {\bibfield  {journal}
  {\bibinfo  {journal} {Phys. Rev. B}\ }\textbf {\bibinfo {volume} {74}},\
  \bibinfo {pages} {125106} (\bibinfo {year} {2006})}\BibitemShut {NoStop}%
\bibitem [{\citenamefont {Casula}\ \emph
  {et~al.}(2012{\natexlab{b}})\citenamefont {Casula}, \citenamefont {Rubtsov},\
  and\ \citenamefont {Biermann}}]{Casula2012a}%
  \BibitemOpen
  \bibfield  {author} {\bibinfo {author} {\bibfnamefont {M.}~\bibnamefont
  {Casula}}, \bibinfo {author} {\bibfnamefont {A.}~\bibnamefont {Rubtsov}},\
  and\ \bibinfo {author} {\bibfnamefont {S.}~\bibnamefont {Biermann}},\ }\href
  {https://doi.org/10.1103/PhysRevB.85.035115} {\bibfield  {journal} {\bibinfo
  {journal} {Phys. Rev. B}\ }\textbf {\bibinfo {volume} {85}},\ \bibinfo
  {pages} {035115} (\bibinfo {year} {2012}{\natexlab{b}})}\BibitemShut
  {NoStop}%
\bibitem [{\citenamefont {Reining}(2018)}]{Reining2018}%
  \BibitemOpen
  \bibfield  {author} {\bibinfo {author} {\bibfnamefont {L.}~\bibnamefont
  {Reining}},\ }\href {https://doi.org/https://doi.org/10.1002/wcms.1344}
  {\bibfield  {journal} {\bibinfo  {journal} {WIREs Computational Molecular
  Science}\ }\textbf {\bibinfo {volume} {8}},\ \bibinfo {pages} {e1344}
  (\bibinfo {year} {2018})},\ \Eprint
  {https://arxiv.org/abs/https://wires.onlinelibrary.wiley.com/doi/pdf/10.1002/wcms.1344}
  {https://wires.onlinelibrary.wiley.com/doi/pdf/10.1002/wcms.1344}
  \BibitemShut {NoStop}%
\bibitem [{\citenamefont {Backes}\ \emph {et~al.}(2022)\citenamefont {Backes},
  \citenamefont {Sim},\ and\ \citenamefont {Biermann}}]{Backes2022}%
  \BibitemOpen
  \bibfield  {author} {\bibinfo {author} {\bibfnamefont {S.}~\bibnamefont
  {Backes}}, \bibinfo {author} {\bibfnamefont {J.-H.}\ \bibnamefont {Sim}},\
  and\ \bibinfo {author} {\bibfnamefont {S.}~\bibnamefont {Biermann}},\ }\href
  {https://doi.org/10.1103/PhysRevB.105.245115} {\bibfield  {journal} {\bibinfo
   {journal} {Phys. Rev. B}\ }\textbf {\bibinfo {volume} {105}},\ \bibinfo
  {pages} {245115} (\bibinfo {year} {2022})}\BibitemShut {NoStop}%
\bibitem [{\citenamefont {Nakamura}\ \emph {et~al.}(2016)\citenamefont
  {Nakamura}, \citenamefont {Nohara}, \citenamefont {Yosimoto},\ and\
  \citenamefont {Nomura}}]{Nakamura2016}%
  \BibitemOpen
  \bibfield  {author} {\bibinfo {author} {\bibfnamefont {K.}~\bibnamefont
  {Nakamura}}, \bibinfo {author} {\bibfnamefont {Y.}~\bibnamefont {Nohara}},
  \bibinfo {author} {\bibfnamefont {Y.}~\bibnamefont {Yosimoto}},\ and\
  \bibinfo {author} {\bibfnamefont {Y.}~\bibnamefont {Nomura}},\ }\href
  {https://doi.org/10.1103/PhysRevB.93.085124} {\bibfield  {journal} {\bibinfo
  {journal} {Phys. Rev. B}\ }\textbf {\bibinfo {volume} {93}},\ \bibinfo
  {pages} {085124} (\bibinfo {year} {2016})}\BibitemShut {NoStop}%
\bibitem [{\citenamefont {Ali}\ \emph {et~al.}(2019)\citenamefont {Ali},
  \citenamefont {Reddy},\ and\ \citenamefont {Singh}}]{Ali2019}%
  \BibitemOpen
  \bibfield  {author} {\bibinfo {author} {\bibfnamefont {A.}~\bibnamefont
  {Ali}}, \bibinfo {author} {\bibfnamefont {B.~H.}\ \bibnamefont {Reddy}},\
  and\ \bibinfo {author} {\bibfnamefont {R.~S.}\ \bibnamefont {Singh}},\ }\href
  {https://doi.org/10.1063/1.5113228} {\bibfield  {journal} {\bibinfo
  {journal} {AIP Conference Proceedings}\ }\textbf {\bibinfo {volume} {2115}},\
  \bibinfo {pages} {030389} (\bibinfo {year} {2019})},\ \Eprint
  {https://arxiv.org/abs/https://aip.scitation.org/doi/pdf/10.1063/1.5113228}
  {https://aip.scitation.org/doi/pdf/10.1063/1.5113228} \BibitemShut {NoStop}%
\bibitem [{\citenamefont {Wadati}\ \emph {et~al.}(2014)\citenamefont {Wadati},
  \citenamefont {Mravlje}, \citenamefont {Yoshimatsu}, \citenamefont
  {Kumigashira}, \citenamefont {Oshima}, \citenamefont {Sugiyama},
  \citenamefont {Ikenaga}, \citenamefont {Fujimori}, \citenamefont {Georges},
  \citenamefont {Radetinac}, \citenamefont {Takahashi}, \citenamefont
  {Kawasaki},\ and\ \citenamefont {Tokura}}]{Wadati2014}%
  \BibitemOpen
  \bibfield  {author} {\bibinfo {author} {\bibfnamefont {H.}~\bibnamefont
  {Wadati}}, \bibinfo {author} {\bibfnamefont {J.}~\bibnamefont {Mravlje}},
  \bibinfo {author} {\bibfnamefont {K.}~\bibnamefont {Yoshimatsu}}, \bibinfo
  {author} {\bibfnamefont {H.}~\bibnamefont {Kumigashira}}, \bibinfo {author}
  {\bibfnamefont {M.}~\bibnamefont {Oshima}}, \bibinfo {author} {\bibfnamefont
  {T.}~\bibnamefont {Sugiyama}}, \bibinfo {author} {\bibfnamefont
  {E.}~\bibnamefont {Ikenaga}}, \bibinfo {author} {\bibfnamefont
  {A.}~\bibnamefont {Fujimori}}, \bibinfo {author} {\bibfnamefont
  {A.}~\bibnamefont {Georges}}, \bibinfo {author} {\bibfnamefont
  {A.}~\bibnamefont {Radetinac}}, \bibinfo {author} {\bibfnamefont {K.~S.}\
  \bibnamefont {Takahashi}}, \bibinfo {author} {\bibfnamefont {M.}~\bibnamefont
  {Kawasaki}},\ and\ \bibinfo {author} {\bibfnamefont {Y.}~\bibnamefont
  {Tokura}},\ }\href {https://doi.org/10.1103/PhysRevB.90.205131} {\bibfield
  {journal} {\bibinfo  {journal} {Phys. Rev. B}\ }\textbf {\bibinfo {volume}
  {90}},\ \bibinfo {pages} {205131} (\bibinfo {year} {2014})}\BibitemShut
  {NoStop}%
\bibitem [{\citenamefont {Radetinac}\ \emph {et~al.}(2016)\citenamefont
  {Radetinac}, \citenamefont {Zimmermann}, \citenamefont {Hoyer}, \citenamefont
  {Zhang}, \citenamefont {Komissinskiy},\ and\ \citenamefont
  {Alff}}]{Radetinac2016}%
  \BibitemOpen
  \bibfield  {author} {\bibinfo {author} {\bibfnamefont {A.}~\bibnamefont
  {Radetinac}}, \bibinfo {author} {\bibfnamefont {J.}~\bibnamefont
  {Zimmermann}}, \bibinfo {author} {\bibfnamefont {K.}~\bibnamefont {Hoyer}},
  \bibinfo {author} {\bibfnamefont {H.}~\bibnamefont {Zhang}}, \bibinfo
  {author} {\bibfnamefont {P.}~\bibnamefont {Komissinskiy}},\ and\ \bibinfo
  {author} {\bibfnamefont {L.}~\bibnamefont {Alff}},\ }\href
  {https://doi.org/10.1063/1.4940969} {\bibfield  {journal} {\bibinfo
  {journal} {Journal of Applied Physics}\ }\textbf {\bibinfo {volume} {119}},\
  \bibinfo {pages} {055302} (\bibinfo {year} {2016})},\ \Eprint
  {https://arxiv.org/abs/https://doi.org/10.1063/1.4940969}
  {https://doi.org/10.1063/1.4940969} \BibitemShut {NoStop}%
\bibitem [{\citenamefont {Chen}\ \emph {et~al.}(2022)\citenamefont {Chen},
  \citenamefont {Petocchi},\ and\ \citenamefont {Werner}}]{Chen2022}%
  \BibitemOpen
  \bibfield  {author} {\bibinfo {author} {\bibfnamefont {J.}~\bibnamefont
  {Chen}}, \bibinfo {author} {\bibfnamefont {F.}~\bibnamefont {Petocchi}},\
  and\ \bibinfo {author} {\bibfnamefont {P.}~\bibnamefont {Werner}},\ }\href
  {https://doi.org/10.1103/PhysRevB.105.085102} {\bibfield  {journal} {\bibinfo
   {journal} {Phys. Rev. B}\ }\textbf {\bibinfo {volume} {105}},\ \bibinfo
  {pages} {085102} (\bibinfo {year} {2022})}\BibitemShut {NoStop}%
\bibitem [{\citenamefont {Su}\ \emph {et~al.}(2023)\citenamefont {Su},
  \citenamefont {Ruotsalainen}, \citenamefont {Nicolaou}, \citenamefont
  {Gatti},\ and\ \citenamefont {Gloter}}]{Gloter2023}%
  \BibitemOpen
  \bibfield  {author} {\bibinfo {author} {\bibfnamefont {C.-P.}\ \bibnamefont
  {Su}}, \bibinfo {author} {\bibfnamefont {K.}~\bibnamefont {Ruotsalainen}},
  \bibinfo {author} {\bibfnamefont {A.}~\bibnamefont {Nicolaou}}, \bibinfo
  {author} {\bibfnamefont {M.}~\bibnamefont {Gatti}},\ and\ \bibinfo {author}
  {\bibfnamefont {A.}~\bibnamefont {Gloter}},\ }\href
  {https://doi.org/https://doi.org/10.1002/adom.202202415} {\bibfield
  {journal} {\bibinfo  {journal} {Advanced Optical Materials}\ }\textbf
  {\bibinfo {volume} {n/a}},\ \bibinfo {pages} {2202415} (\bibinfo {year}
  {2023})},\ \Eprint
  {https://arxiv.org/abs/https://onlinelibrary.wiley.com/doi/pdf/10.1002/adom.202202415}
  {https://onlinelibrary.wiley.com/doi/pdf/10.1002/adom.202202415} \BibitemShut
  {NoStop}%
\end{thebibliography}%


\begin{thebibliography}{13}%
\makeatletter
\providecommand \@ifxundefined [1]{%
 \@ifx{#1\undefined}
}%
\providecommand \@ifnum [1]{%
 \ifnum #1\expandafter \@firstoftwo
 \else \expandafter \@secondoftwo
 \fi
}%
\providecommand \@ifx [1]{%
 \ifx #1\expandafter \@firstoftwo
 \else \expandafter \@secondoftwo
 \fi
}%
\providecommand \natexlab [1]{#1}%
\providecommand \enquote  [1]{``#1''}%
\providecommand \bibnamefont  [1]{#1}%
\providecommand \bibfnamefont [1]{#1}%
\providecommand \citenamefont [1]{#1}%
\providecommand \href@noop [0]{\@secondoftwo}%
\providecommand \href [0]{\begingroup \@sanitize@url \@href}%
\providecommand \@href[1]{\@@startlink{#1}\@@href}%
\providecommand \@@href[1]{\endgroup#1\@@endlink}%
\providecommand \@sanitize@url [0]{\catcode `\\12\catcode `\$12\catcode
  `\&12\catcode `\#12\catcode `\^12\catcode `\_12\catcode `\%12\relax}%
\providecommand \@@startlink[1]{}%
\providecommand \@@endlink[0]{}%
\providecommand \url  [0]{\begingroup\@sanitize@url \@url }%
\providecommand \@url [1]{\endgroup\@href {#1}{\urlprefix }}%
\providecommand \urlprefix  [0]{URL }%
\providecommand \Eprint [0]{\href }%
\providecommand \doibase [0]{https://doi.org/}%
\providecommand \selectlanguage [0]{\@gobble}%
\providecommand \bibinfo  [0]{\@secondoftwo}%
\providecommand \bibfield  [0]{\@secondoftwo}%
\providecommand \translation [1]{[#1]}%
\providecommand \BibitemOpen [0]{}%
\providecommand \bibitemStop [0]{}%
\providecommand \bibitemNoStop [0]{.\EOS\space}%
\providecommand \EOS [0]{\spacefactor3000\relax}%
\providecommand \BibitemShut  [1]{\csname bibitem#1\endcsname}%
\let\auto@bib@innerbib\@empty
\bibitem [{\citenamefont {P}\ \emph {et~al.}(2001)\citenamefont {P},
  \citenamefont {G{ K }H}, \citenamefont {D},\ and\ \citenamefont
  {J}}]{Wien2k}%
  \BibitemOpen
  \bibfield  {author} {\bibinfo {author} {\bibfnamefont {B.}~\bibnamefont {P}},
  \bibinfo {author} {\bibfnamefont {S.~M.}\ \bibnamefont {G{ K }H}}, \bibinfo
  {author} {\bibfnamefont {K.}~\bibnamefont {D}},\ and\ \bibinfo {author}
  {\bibfnamefont {L.}~\bibnamefont {J}},\ }\href@noop {} {\bibfield  {journal}
  {\bibinfo  {journal} {Karlheinz Schwarz, Techn. Universit\"at Wien, Austria}\
  } (\bibinfo {year} {2001})}\BibitemShut {NoStop}%
\bibitem [{\citenamefont {Jiang}\ \emph {et~al.}(2012)\citenamefont {Jiang},
  \citenamefont {G\'omez-Abal}, \citenamefont {Li}, \citenamefont
  {Meisenbichler}, \citenamefont {Ambrosch-Draxl}, ,\ and\ \citenamefont
  {Scheffler}}]{fhigap}%
  \BibitemOpen
  \bibfield  {author} {\bibinfo {author} {\bibfnamefont {H.}~\bibnamefont
  {Jiang}}, \bibinfo {author} {\bibfnamefont {R.~I.}\ \bibnamefont
  {G\'omez-Abal}}, \bibinfo {author} {\bibfnamefont {X.}~\bibnamefont {Li}},
  \bibinfo {author} {\bibfnamefont {C.}~\bibnamefont {Meisenbichler}}, \bibinfo
  {author} {\bibfnamefont {C.}~\bibnamefont {Ambrosch-Draxl}}, ,\ and\ \bibinfo
  {author} {\bibfnamefont {M.}~\bibnamefont {Scheffler}},\ }\href@noop {}
  {\bibfield  {journal} {\bibinfo  {journal} {Computer Phys. Commun.,184, 348}\
  }\textbf {\bibinfo {volume} {184}},\ \bibinfo {pages} {348} (\bibinfo {year}
  {2012})}\BibitemShut {NoStop}%
\bibitem [{\citenamefont {Gaenko}\ \emph {et~al.}(2017)\citenamefont {Gaenko},
  \citenamefont {Antipov}, \citenamefont {Carcassi}, \citenamefont {Chen},
  \citenamefont {Chen}, \citenamefont {Dong}, \citenamefont {Gamper},
  \citenamefont {Gukelberger}, \citenamefont {Igarashi}, \citenamefont
  {Iskakov}, \citenamefont {Könz}, \citenamefont {LeBlanc}, \citenamefont
  {Levy}, \citenamefont {Ma}, \citenamefont {Paki}, \citenamefont {Shinaoka},
  \citenamefont {Todo}, \citenamefont {Troyer},\ and\ \citenamefont
  {Gull}}]{ALPSCore}%
  \BibitemOpen
  \bibfield  {author} {\bibinfo {author} {\bibfnamefont {A.}~\bibnamefont
  {Gaenko}}, \bibinfo {author} {\bibfnamefont {A.}~\bibnamefont {Antipov}},
  \bibinfo {author} {\bibfnamefont {G.}~\bibnamefont {Carcassi}}, \bibinfo
  {author} {\bibfnamefont {T.}~\bibnamefont {Chen}}, \bibinfo {author}
  {\bibfnamefont {X.}~\bibnamefont {Chen}}, \bibinfo {author} {\bibfnamefont
  {Q.}~\bibnamefont {Dong}}, \bibinfo {author} {\bibfnamefont {L.}~\bibnamefont
  {Gamper}}, \bibinfo {author} {\bibfnamefont {J.}~\bibnamefont {Gukelberger}},
  \bibinfo {author} {\bibfnamefont {R.}~\bibnamefont {Igarashi}}, \bibinfo
  {author} {\bibfnamefont {S.}~\bibnamefont {Iskakov}}, \bibinfo {author}
  {\bibfnamefont {M.}~\bibnamefont {Könz}}, \bibinfo {author} {\bibfnamefont
  {J.}~\bibnamefont {LeBlanc}}, \bibinfo {author} {\bibfnamefont
  {R.}~\bibnamefont {Levy}}, \bibinfo {author} {\bibfnamefont {P.}~\bibnamefont
  {Ma}}, \bibinfo {author} {\bibfnamefont {J.}~\bibnamefont {Paki}}, \bibinfo
  {author} {\bibfnamefont {H.}~\bibnamefont {Shinaoka}}, \bibinfo {author}
  {\bibfnamefont {S.}~\bibnamefont {Todo}}, \bibinfo {author} {\bibfnamefont
  {M.}~\bibnamefont {Troyer}},\ and\ \bibinfo {author} {\bibfnamefont
  {E.}~\bibnamefont {Gull}},\ }\href
  {https://doi.org/https://doi.org/10.1016/j.cpc.2016.12.009} {\bibfield
  {journal} {\bibinfo  {journal} {Computer Physics Communications}\ }\textbf
  {\bibinfo {volume} {213}},\ \bibinfo {pages} {235 } (\bibinfo {year}
  {2017})}\BibitemShut {NoStop}%
\bibitem [{\citenamefont {Levy}\ \emph {et~al.}(2017)\citenamefont {Levy},
  \citenamefont {LeBlanc},\ and\ \citenamefont {Gull}}]{levi_maxent}%
  \BibitemOpen
  \bibfield  {author} {\bibinfo {author} {\bibfnamefont {R.}~\bibnamefont
  {Levy}}, \bibinfo {author} {\bibfnamefont {J.}~\bibnamefont {LeBlanc}},\ and\
  \bibinfo {author} {\bibfnamefont {E.}~\bibnamefont {Gull}},\ }\href
  {https://doi.org/https://doi.org/10.1016/j.cpc.2017.01.018} {\bibfield
  {journal} {\bibinfo  {journal} {Computer Physics Communications}\ }\textbf
  {\bibinfo {volume} {215}},\ \bibinfo {pages} {149 } (\bibinfo {year}
  {2017})}\BibitemShut {NoStop}%
\bibitem [{\citenamefont {Lange}(1998)}]{Lange_1998}%
  \BibitemOpen
  \bibfield  {author} {\bibinfo {author} {\bibfnamefont {E.}~\bibnamefont
  {Lange}},\ }\href {https://doi.org/10.1142/s0217984998001050} {\bibfield
  {journal} {\bibinfo  {journal} {Modern Physics Letters B}\ }\textbf {\bibinfo
  {volume} {12}},\ \bibinfo {pages} {915} (\bibinfo {year} {1998})}\BibitemShut
  {NoStop}%
\bibitem [{\citenamefont {Bulla}\ and\ \citenamefont
  {Potthoff}(2000)}]{Bulla_2000}%
  \BibitemOpen
  \bibfield  {author} {\bibinfo {author} {\bibfnamefont {R.}~\bibnamefont
  {Bulla}}\ and\ \bibinfo {author} {\bibfnamefont {M.}~\bibnamefont
  {Potthoff}},\ }\href {https://doi.org/10.1007/s100510050030} {\bibfield
  {journal} {\bibinfo  {journal} {The European Physical Journal B}\ }\textbf
  {\bibinfo {volume} {13}},\ \bibinfo {pages} {257} (\bibinfo {year}
  {2000})}\BibitemShut {NoStop}%
\bibitem [{\citenamefont {Potthoff}(2001)}]{Potthoff2001}%
  \BibitemOpen
  \bibfield  {author} {\bibinfo {author} {\bibfnamefont {M.}~\bibnamefont
  {Potthoff}},\ }\href {https://doi.org/10.1103/PhysRevB.64.165114} {\bibfield
  {journal} {\bibinfo  {journal} {Phys. Rev. B}\ }\textbf {\bibinfo {volume}
  {64}},\ \bibinfo {pages} {165114} (\bibinfo {year} {2001})}\BibitemShut
  {NoStop}%
\bibitem [{\citenamefont {Boehnke}\ \emph {et~al.}(2016)\citenamefont
  {Boehnke}, \citenamefont {Nilsson}, \citenamefont {Aryasetiawan},\ and\
  \citenamefont {Werner}}]{Boehnke2016}%
  \BibitemOpen
  \bibfield  {author} {\bibinfo {author} {\bibfnamefont {L.}~\bibnamefont
  {Boehnke}}, \bibinfo {author} {\bibfnamefont {F.}~\bibnamefont {Nilsson}},
  \bibinfo {author} {\bibfnamefont {F.}~\bibnamefont {Aryasetiawan}},\ and\
  \bibinfo {author} {\bibfnamefont {P.}~\bibnamefont {Werner}},\ }\href
  {https://doi.org/10.1103/PhysRevB.94.201106} {\bibfield  {journal} {\bibinfo
  {journal} {Phys. Rev. B}\ }\textbf {\bibinfo {volume} {94}},\ \bibinfo
  {pages} {201106(R)} (\bibinfo {year} {2016})}\BibitemShut {NoStop}%
\bibitem [{\citenamefont {Nilsson}\ \emph {et~al.}(2017)\citenamefont
  {Nilsson}, \citenamefont {Boehnke}, \citenamefont {Werner},\ and\
  \citenamefont {Aryasetiawan}}]{Nilsson2017}%
  \BibitemOpen
  \bibfield  {author} {\bibinfo {author} {\bibfnamefont {F.}~\bibnamefont
  {Nilsson}}, \bibinfo {author} {\bibfnamefont {L.}~\bibnamefont {Boehnke}},
  \bibinfo {author} {\bibfnamefont {P.}~\bibnamefont {Werner}},\ and\ \bibinfo
  {author} {\bibfnamefont {F.}~\bibnamefont {Aryasetiawan}},\ }\href
  {https://doi.org/10.1103/PhysRevMaterials.1.043803} {\bibfield  {journal}
  {\bibinfo  {journal} {Phys. Rev. Materials}\ }\textbf {\bibinfo {volume}
  {1}},\ \bibinfo {pages} {043803} (\bibinfo {year} {2017})}\BibitemShut
  {NoStop}%
\bibitem [{\citenamefont {Evtushinsky}\ \emph {et~al.}(2016)\citenamefont
  {Evtushinsky}, \citenamefont {Aichhorn}, \citenamefont {Y.~Sassa},
  \citenamefont {Maletz}, \citenamefont {T.Wolf}, \citenamefont {N.Yaresko},
  \citenamefont {Biermann}, \citenamefont {Borisenko},\ and\ \citenamefont
  {B.Buchner}}]{Evtushinsky2016}%
  \BibitemOpen
  \bibfield  {author} {\bibinfo {author} {\bibfnamefont {D.~V.}\ \bibnamefont
  {Evtushinsky}}, \bibinfo {author} {\bibfnamefont {M.}~\bibnamefont
  {Aichhorn}}, \bibinfo {author} {\bibfnamefont {Z.-H.~L.}\ \bibnamefont
  {Y.~Sassa}}, \bibinfo {author} {\bibfnamefont {J.}~\bibnamefont {Maletz}},
  \bibinfo {author} {\bibnamefont {T.Wolf}}, \bibinfo {author} {\bibfnamefont
  {A.}~\bibnamefont {N.Yaresko}}, \bibinfo {author} {\bibfnamefont
  {S.}~\bibnamefont {Biermann}}, \bibinfo {author} {\bibfnamefont {S.~V.}\
  \bibnamefont {Borisenko}},\ and\ \bibinfo {author} {\bibnamefont
  {B.Buchner}},\ }\href {https://arxiv.org/abs/1612.02313} {\bibfield
  {journal} {\bibinfo  {journal} {arxiv.org}\ }\textbf {\bibinfo {volume}
  {arXiv:1612.02313}} (\bibinfo {year} {2016})}\BibitemShut {NoStop}%
\bibitem [{\citenamefont {Tomczak}\ \emph {et~al.}(2014)\citenamefont
  {Tomczak}, \citenamefont {Casula}, \citenamefont {Miyake},\ and\
  \citenamefont {Biermann}}]{Tomczak2014}%
  \BibitemOpen
  \bibfield  {author} {\bibinfo {author} {\bibfnamefont {J.~M.}\ \bibnamefont
  {Tomczak}}, \bibinfo {author} {\bibfnamefont {M.}~\bibnamefont {Casula}},
  \bibinfo {author} {\bibfnamefont {T.}~\bibnamefont {Miyake}},\ and\ \bibinfo
  {author} {\bibfnamefont {S.}~\bibnamefont {Biermann}},\ }\href
  {https://doi.org/10.1103/PhysRevB.90.165138} {\bibfield  {journal} {\bibinfo
  {journal} {Phys. Rev. B}\ }\textbf {\bibinfo {volume} {90}},\ \bibinfo
  {pages} {165138} (\bibinfo {year} {2014})}\BibitemShut {NoStop}%
\bibitem [{\citenamefont {Werner}\ and\ \citenamefont
  {Millis}(2010)}]{Werner2010}%
  \BibitemOpen
  \bibfield  {author} {\bibinfo {author} {\bibfnamefont {P.}~\bibnamefont
  {Werner}}\ and\ \bibinfo {author} {\bibfnamefont {A.~J.}\ \bibnamefont
  {Millis}},\ }\href {https://doi.org/10.1103/PhysRevLett.104.146401}
  {\bibfield  {journal} {\bibinfo  {journal} {Phys. Rev. Lett.}\ }\textbf
  {\bibinfo {volume} {104}},\ \bibinfo {pages} {146401} (\bibinfo {year}
  {2010})}\BibitemShut {NoStop}%
\bibitem [{\citenamefont {Casula}\ \emph {et~al.}(2012)\citenamefont {Casula},
  \citenamefont {Rubtsov},\ and\ \citenamefont {Biermann}}]{Casula2012a}%
  \BibitemOpen
  \bibfield  {author} {\bibinfo {author} {\bibfnamefont {M.}~\bibnamefont
  {Casula}}, \bibinfo {author} {\bibfnamefont {A.}~\bibnamefont {Rubtsov}},\
  and\ \bibinfo {author} {\bibfnamefont {S.}~\bibnamefont {Biermann}},\ }\href
  {https://doi.org/10.1103/PhysRevB.85.035115} {\bibfield  {journal} {\bibinfo
  {journal} {Phys. Rev. B}\ }\textbf {\bibinfo {volume} {85}},\ \bibinfo
  {pages} {035115} (\bibinfo {year} {2012})}\BibitemShut {NoStop}%
\end{thebibliography}%

 \end{document}


\author{Steffen Backes$^{1,2,3}$}
\email[]{steffen-backes@g.ecc.u-tokyo.ac.jp}
\author{Hong Jiang$^4$}
\author{Silke Biermann$^{3,5,6,7}$}
\affiliation{$^1$Research Center for Advanced Science and Technology, University of Tokyo, Komaba, Tokyo 153-8904, Japan}
\affiliation{$^2$Center for Emergent Matter Science, RIKEN, Wako, Saitama 351-0198, Japan}
\affiliation{$^3$CPHT, CNRS, École polytechnique, Institut Polytechnique de Paris, 91120 Palaiseau, France}
\affiliation{$^4$College of Chemistry and Molecular Engineering, Peking University, China}
\affiliation{$^5$Coll\`ege de France, 11 place Marcelin Berthelot, 75005 Paris, France}
\affiliation{$^6$European Theoretical Spectroscopy Facility, 91128 Palaiseau, France, Europe}
\affiliation{$^7$Department of Physics, Division of Mathematical Physics, Lund University, Professorsgatan 1, 22363 Lund, Sweden}

\date{\today}


\
\vspace{0.5cm}
\

\title{Diagnostics for plasmon satellites and Hubbard bands in transition metal oxides - Supplementary Material}

\maketitle

\section{Computational details}
For the GW+EDMFT cycle we start with a well converged DFT calculation from Wien2K~\cite{Wien2k},
and perform a constrained Random-Phase-Approximation and a $G_0W_0$ calculation, 
as implemented in the FHI-gap Code\cite{fhigap}, to obtain the effective impurity interaction $U(\omega)$
and the Selfenergy $\Sigma_{GW}(k,i\omega_n)$, projected onto the $t_{2g}$ orbitals
of either {\svo} or {\smo}, using a maximally localized Wannier basis.
The cRPA and GW calculation were performed on a $8\times8 \times 8$ k-mesh,
and the resulting $U(\omega)$ and $\Sigma_{GW}(k,i\omega_n)$ are then interpolated by cubic interpolation
onto a dense $30\times 30 \times 30$ k-mesh, which serves as the input for the 
selfconsistent EDMFT calculation. The impurity model is solved within the 
continuous-time Quantum Monte-Carlo method in the hybridization expansion,
as implemented in the ALPS package\cite{ALPSCore} at inverse temperature $\beta=40~$1/eV,
including the frequency dependence of the monopole term $F_0(\omega)$ of the effective interaction.
The analytical continuation from the imaginary to the real frequency axis is done using 
a combination of Pad\'e approximants and the 
Maximum Entropy code from Ref.~\cite{levi_maxent}, where we added plasmonic peaks in the default model
of exponentially decaying weight at multiples of the plasma frequency to faciliate
proper continuation.

\begin{figure}[t]
\includegraphics[width=0.5\textwidth]{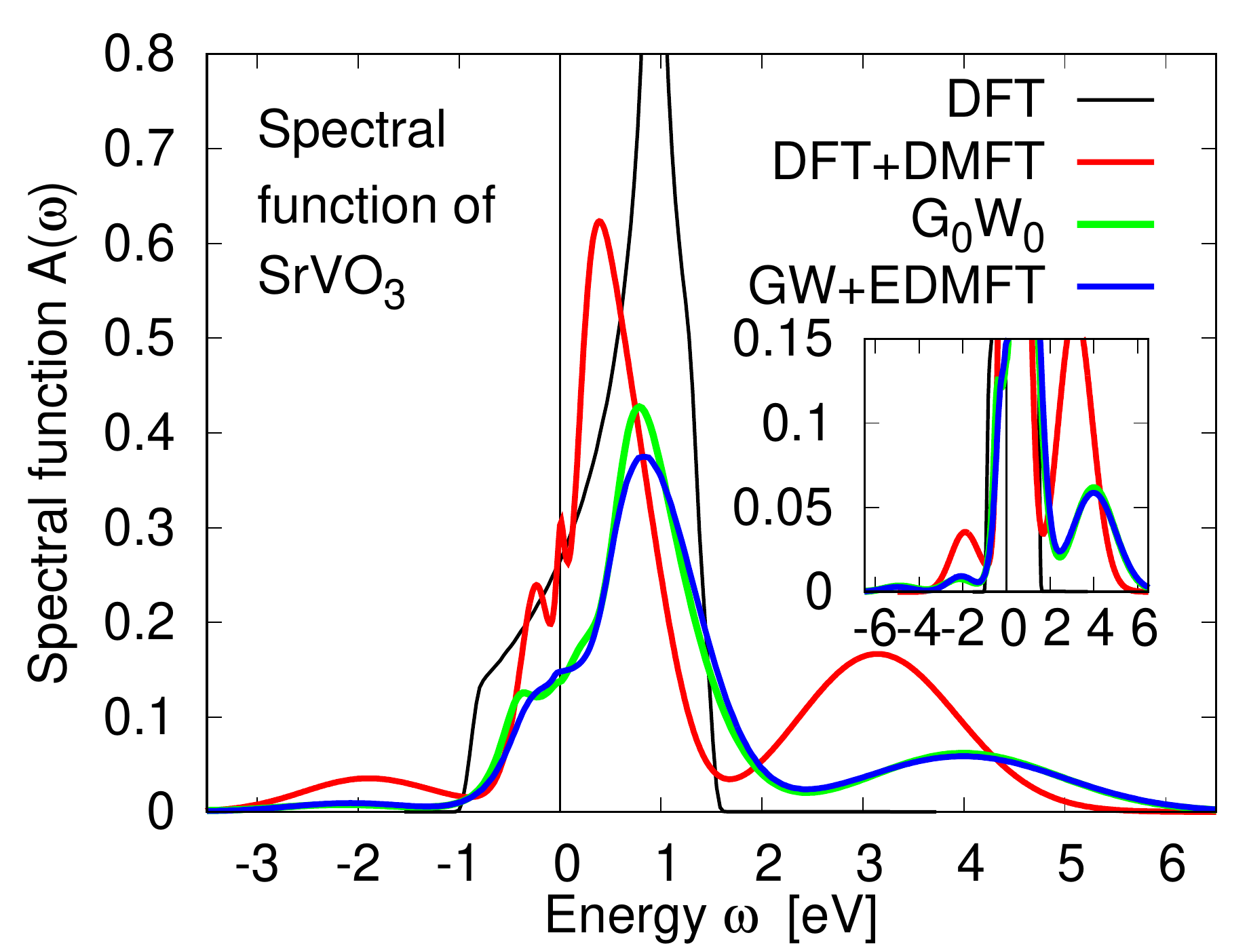} 
\caption{
The spectral function of {\svo}, comparing the Desity Functional Theory (DFT), 
the DFT+DMFT, the $G_0W_0$ approximation and the $GW$+EDMFT result shown in the main text.
The $G_0W_0$ and the $GW$+EDMFT spectral functions are almost identical, indicating
that a fully selfconsistent $GW$+EDMFT calculation will not significantly change the result.
}
\label{fig:srvo3_gwgwdmft_comp}
\end{figure}
The GW+EDMFT calculation was not done fully self-consistently, i.e. the 
nonlocal $GW$ self-energy remained at the $G_0W_0$ level, and the effective interaction $U(i\omega)$
was not updated. 
As discussed in the main text the resulting spectral function from this GW+EDMFT
is very close to the $G_0W_0$ result, as shown in Fig.\ref{fig:srvo3_gwgwdmft_comp}.
This indicates that further self-consistency will have only minor effects and not qualitatively change 
the one-shot result. Therefore, we only considered the 'one-shot' $GW$+EDMFT results, 
which are computationally less demanding.

\section{Local $G_0W_0$ and satellites}
The results shown in Fig. 3 in the main text were obtained by employing a local $G_0W_0$ approximation:
First, the polarization was calculated from the projected local DFT Green's function for the $t_{2g}$ orbitals
as $P_{loc}=G_{loc}G_{loc}$. For the effective ``bare'' interaction the cRPA derived
impurity interaction $U(\omega)$ was used, which was screened by the local polarization 
to obtain the screened local interaction $W_{loc} = U[1-P_{loc}U]^{-1}$ 
(shown in the inset of Fig. 3 in the main text). Then the self-energy was obtained
by the convolution of the local Green's function and screened interaction as $\Sigma=G_{loc} W_{loc}$.

\begin{figure}[t]
\includegraphics[width=0.5\textwidth]{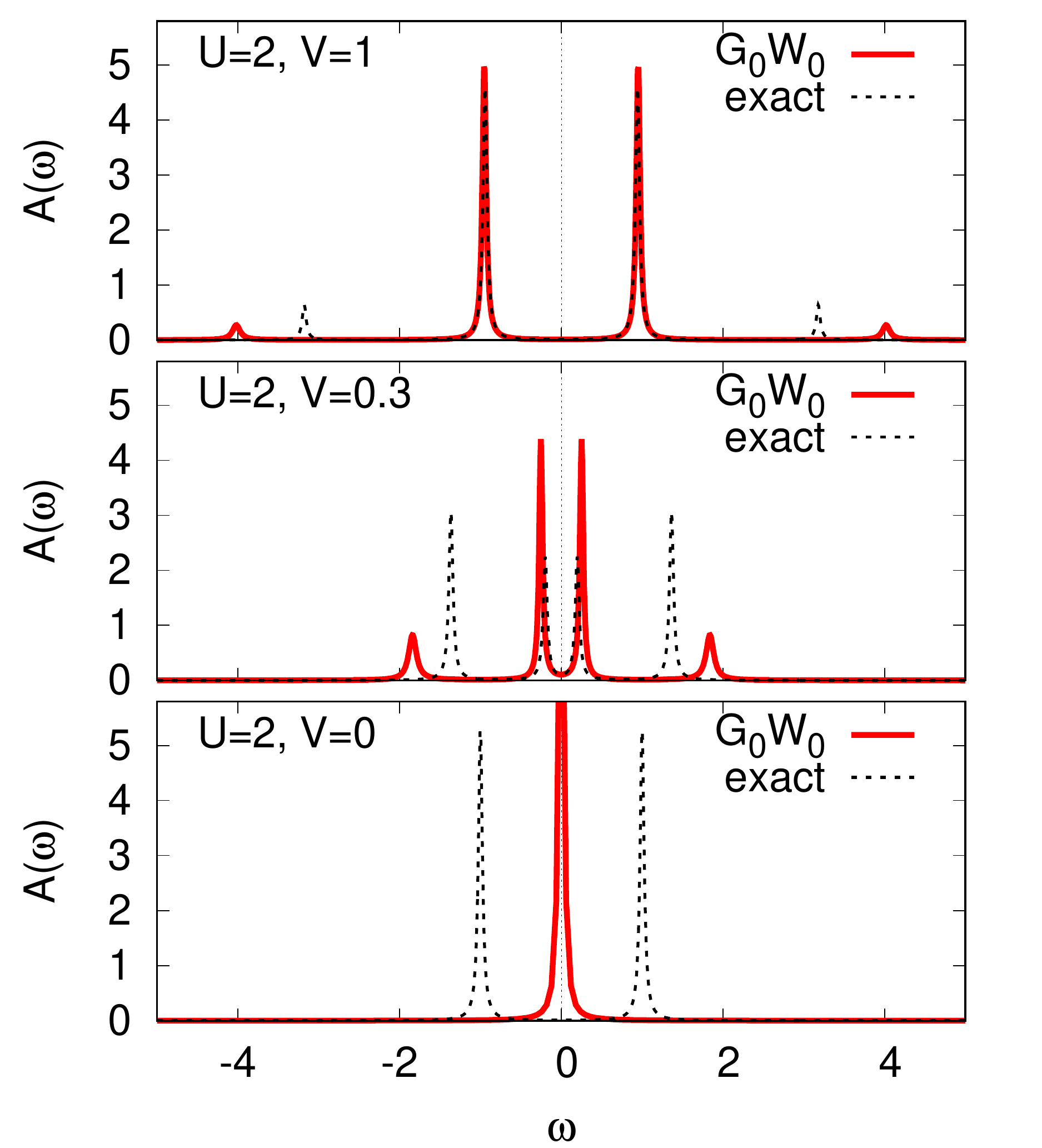} 
\caption{
The ground state spectral function for a DMFT impurity model with one bath site solved
within the $G_0W_0$ approximation for different
values of the hybridization strength $V$ at fixed interaction $U=2$.
Except in the strong interaction limit $V\rightarrow 0$ $G_0W_0$ captures all
qualitative features and correlation satellites, albeit overestimating their 
energetic position.
}
\label{fig:lin_dmft_gw}
\end{figure}

\begin{figure}[t]
\includegraphics[width=0.5\textwidth]{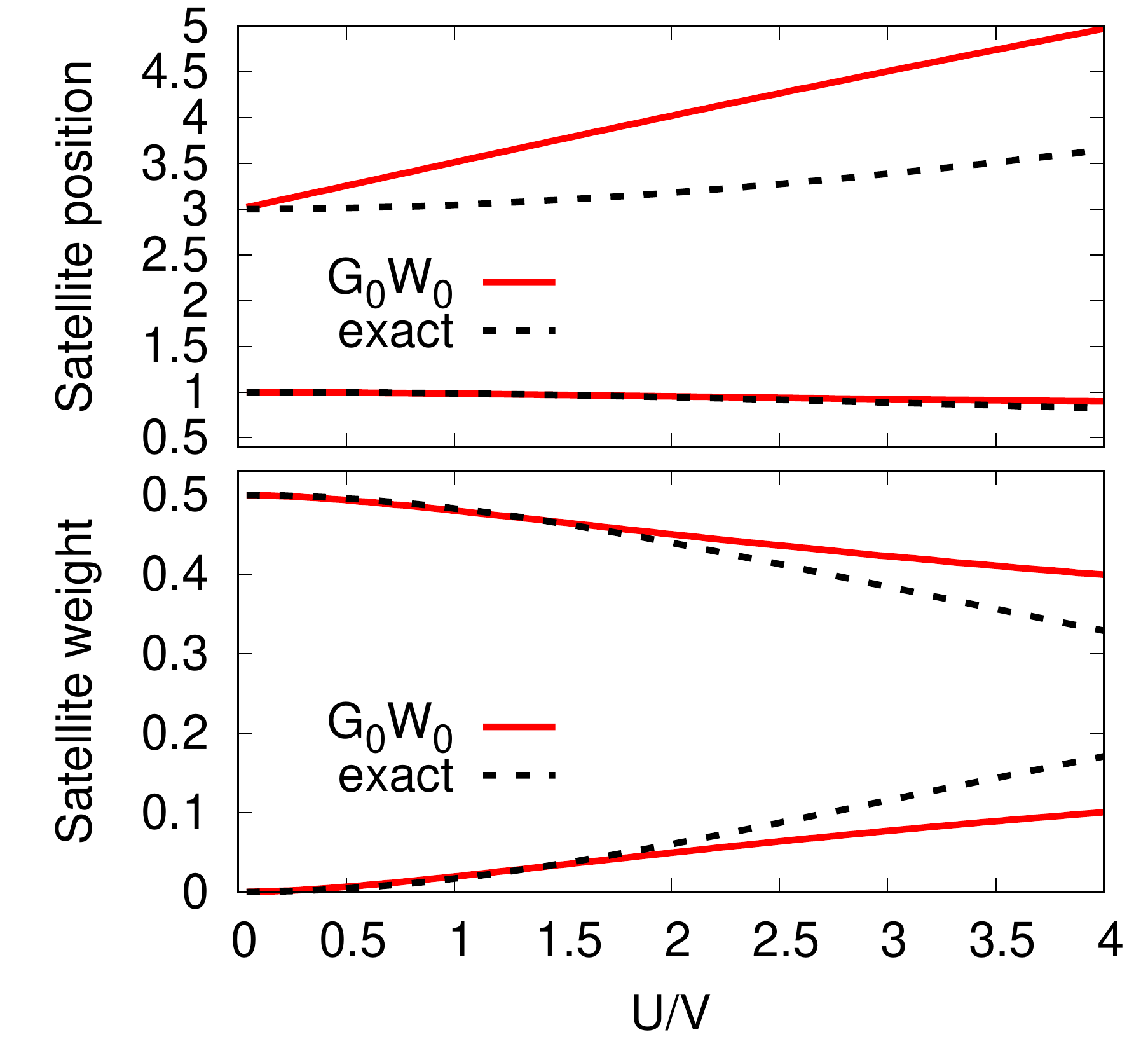} 
\caption{
The satellite positions and weights for the DMFT impurity model with one bath site 
as shown in Fig.\ref{fig:lin_dmft_gw},
as a function of the interaction strength $U$ for fixed hybridization $V=1$.
}
\label{fig:lin_dmft_gw_pos}
\end{figure}

To elucidate the appearance of the Hubbard correlation satellites within $G_0W_0$
one can consider a simplified 'linearized' DMFT impurity problem~\cite{Lange_1998,Bulla_2000,Potthoff2001}
with only one bath site, which can be solved analytically
\begin{align}
 H=Un_{\uparrow}n_{\downarrow} + 
 V\sum_{\sigma} (c^{\dagger}_{\sigma}f_{\sigma} + f^{\dagger}_{\sigma}c_{\sigma})
 -\mu \sum_{\sigma}n_{\sigma},
\end{align}
with $c,c^{\dagger}/f,f^{\dagger}$ the impurity/bath annihilation and creation 
operators, interaction $U$, hybridization strength $V$ and chemical potential $\mu$.
Solving this model exactly and within the $G_0W_0$ approximation yields for the 
impurity self-energy
\begin{align}
 \Sigma_{exact}(z) &= \frac{U^2}{8} \left( \frac{1}{z-3V} + \frac{1}{z+3V}  \right) \\
 \Sigma_{G_0W_0}(z) &= \frac{U^2}{4a} \left( \frac{1}{z-V(2a+1)} + \frac{1}{z+V(2a+1)}  \right) ,
\end{align}
with $a=\sqrt{ 1+ U/(2V) }$. Both results qualitatively agree, but $G_0W_0$ does not capture the correct 
position and weight of the two peaks in $\Sigma$. We note, however, that for
$U/V\rightarrow 0$ the peak position is correctly reproduced in $G_0W_0$, while 
the weight differs by a factor of 2.
In Fig.\ref{fig:lin_dmft_gw} we show the resulting spectral function for different values
of the hybridization $V$ for fixed interaction $U=2$. 
Except in the strong interaction limit $V\rightarrow 0$ $G_0W_0$ captures all
qualitative features and correlation satellites, albeit overestimating their 
energetic position. 
This can be seen explicitly in Fig.\ref{fig:lin_dmft_gw_pos}, where we show the position and weights of the two different
satellites emerging in the spectral function as a function of the interaction strength $U$.
$G_0W_0$ performs reasonably well for small values of the interaction strength $U/V \lesssim 1.5$, 
in particular for the bonding/antibonding states and satellite weights.
The emergence of the correlation satellites in $G_0W_0$, which in this setup
appear as Hubbard satellites, is in fact not surprising, since
a perturbative approach such as  the $G_0W_0$ approximation is
expected to become more accurate in the weakly correlated regime.
If Hubbard satellites in this regime are present, it is
expected that $G_0W_0$ is able to qualitatively capture them,
as shown above.

\section{Separation of Hubbard satellites}
Previously it had been suggested to use the energetic separation
of the unoccupied and occupied satellites, and their dependence on the bandwidth
to distinguish plasmon satellites from Hubbard satellites\cite{Boehnke2016,Nilsson2017}.
In the atomic limit Hubbard satellites are separated by the static onsite interaction $U(0)$,
but for metallic systems a finite bandwidth enhances the separation of the Hubbard satellites\cite{Evtushinsky2016}.
In systems far away from half-filling such as {\svo}, the position of the Hubbard satellites is further complicated
by the breaking of particle-hole symmetry. Already at the DMFT level, the separation of the Hubbard satellites
$\Delta_{Hub}\approx 5$~eV greatly surpasses the value of the static interaction $U \approx 3.5$~eV.

\begin{figure}[t]
\includegraphics[width=0.5\textwidth]{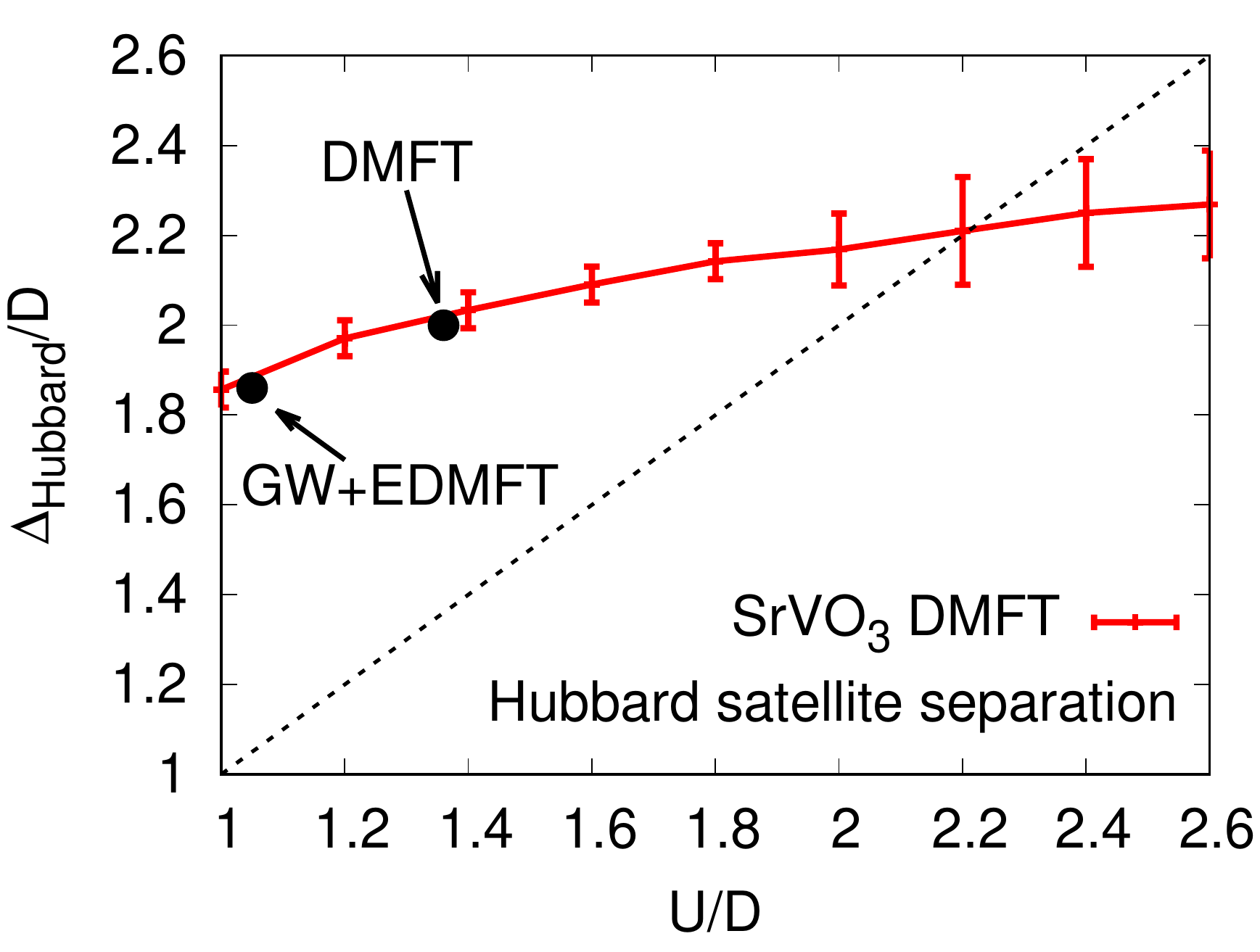} 
\caption{
The separation of the Hubbard satellites in {\svo} within a DMFT approach
for different values of the interaction $U$, relative to the non-interacting bandwidth $D$.
For small interactions the separation is much larger that the value of $U$, 
but becomes smaller for larger interactions. The non-interacting bandwidth for $GW$+EDMFT
is given by the spectral function where the local self-energy has been removed ($\approx 3.25$~eV).
The error bars indicate the uncertainty from analytic continuation.
}
\label{fig:hubbard_satellites_position}
\end{figure}

In Fig.\ref{fig:hubbard_satellites_position} we show the dependence of the Hubbard satellite separation 
in {\svo} on the interaction $U$, obtained from a DMFT calculation.
Up to moderate correlations (for a filling of $n=1/6$) the Hubbard satellite separation exceeds the 
value of $U$ by almost a factor of $2$. On the other hand, for stronger interactions when the quasiparticle 
peak is close to vanishing the situation reverses and the separation becomes smaller than $U$.
For the case of the GW+EDMFT result, which has a larger 'non-interacting' bandwidth $D$ due to the 
removal of the exchange-correlation potential and non-local $GW$ contributions\cite{Tomczak2014},
the Hubbard satellite separation is larger than in standard DMFT, but is well within the expected range,
considering the value of $U(0)/D$ (see label 'GW+EDMFT' in Fig.\ref{fig:hubbard_satellites_position}).
Therefore, the Hubbard satellite separation alone is not a good quantifier to distinguish
plasmonic from Hubbard satellites, as in metallic systems and systems away from half-filling, 
the Hubbard satellite separation significantly differs from the usual $\Delta_{Hub} \sim U$ behavior.

\section{One-orbital model for plasmonic and Hubbard satellites}
\label{sec:model}
In order to distinguish satellites of plasmonic origin and that
arising from strong electronic correlations, we consider a 
simple model that describes the coupling of electrons
to a single bosonic degree of freedom, namely the Hubbard-Holstein model
\begin{align}
 H = & - \sum_{ij}t_{ij} c^{\dagger}_{i\sigma}c_{j\sigma}
  + U_{bare}\sum_{i} n_{i\uparrow}n_{i\downarrow} \nonumber  \\
  &+ \omega_0 \sum_{i} b^{\dagger}_i b_i
  + \lambda\sum_{i} n_i\left(  b^{\dagger}_i + b_i \right) ,
\end{align}
where $t_{ij}$ is the hopping amplitude, $U_{bare}$ is the local instantaneous Coulomb
repulsion, $\omega_0$ is the energy of the bosonic mode (plasma frequency),
generated by the bosonic annihilation and creation
operators $b^{\dagger}_i, b_i$. The coupling strength between the 
electronic charge $n_i$ and the bosonic mode is given by $\lambda$.

Integrating out the bosonic degrees of freedom gives rise to an effective
dynamical interaction $U(\omega)$, where the coupling to
the bosonic mode is now encoded in the frequency dependence
\begin{align}
\mathrm{Re}\, U_{\mathrm{eff}}(\omega) &= U_{bare} - 2\lambda^2\frac{\omega_0}{\omega_0^2 - \omega^2} \\
\mathrm{Im}\, U_{\mathrm{eff}}(\omega) &= -\lambda^2 \pi \left( \delta(\omega-\omega_0) - \delta(\omega+\omega_0) \right).
\label{eq:u_eff}
\end{align}

This model is an extension of the standard Hubbard model, and
it has been shown that this model exhibits plasmonic replicas
of the {\qp} structure at multiples of the 
plasma frequency $\omega_0$, that originate from plasmonic
charge excitations\cite{Werner2010,Casula2012a}. 

\begin{figure}[t]
\includegraphics[width=0.5\textwidth]{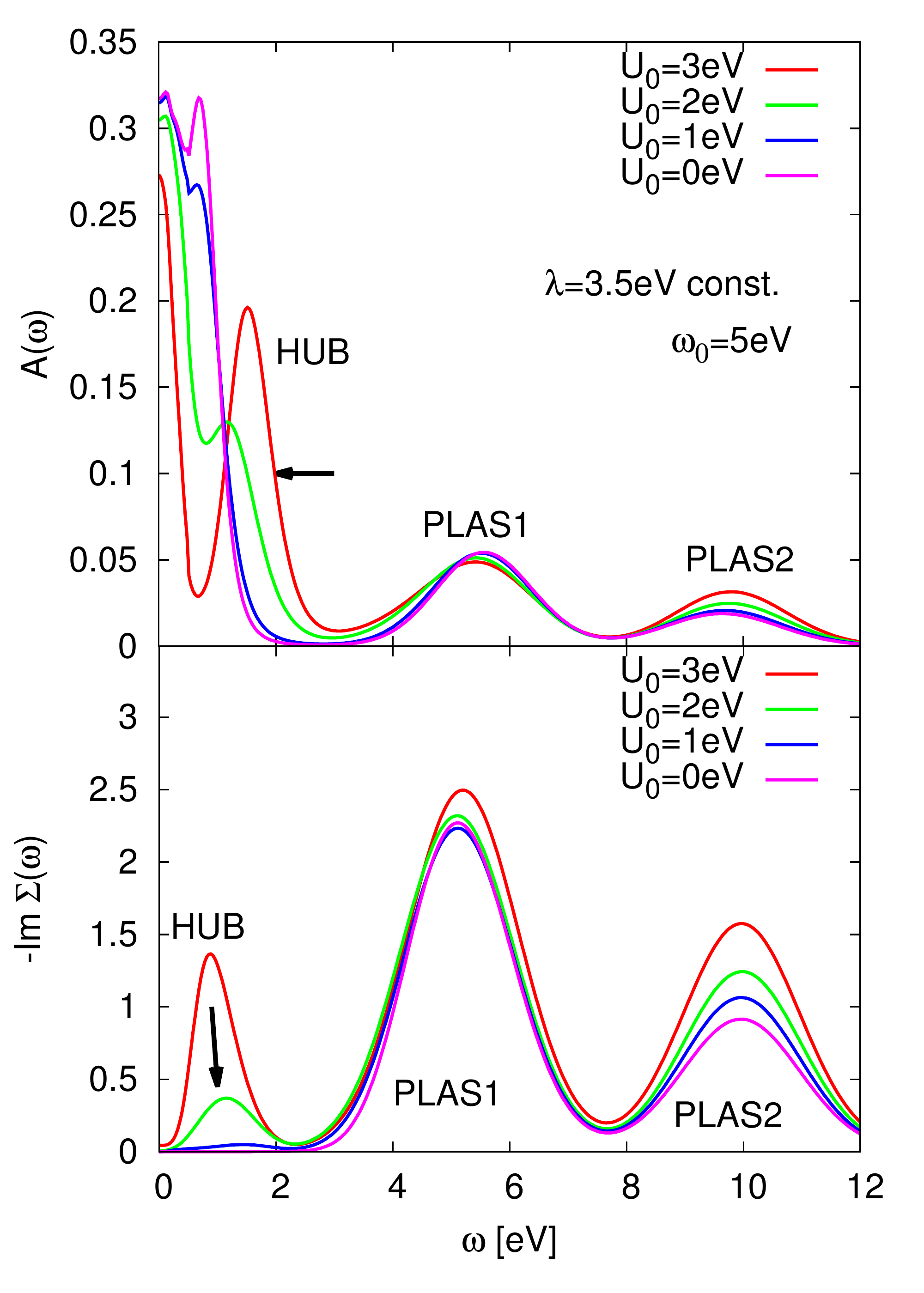} 
\caption{The spectral function and imaginary part of the Selfenergy
for the half-filled Hubbard model with a one boson screening mode. We show only positive 
energies since the spectrum is particle-hole symmetric.
The frequency dependence of the effective interaction $U(\omega)$ has been fixed 
with $U(\infty)-U(0)=5$~eV, but a static shift has been applied in order to reduce
the strength of the interaction but keep the transfer of spectral weight
due to the bosonic coupling constant.
The bosonic energy is $\omega_0=5$~eV, giving rise to plasmonic replica at multiples of $\omega_0$.
We observe a disappearance of the Hubbard satellite as $U(0)$ approaches zero, while the plasmonic
satellites stay mostly unchanged.}
\label{fig:spec_Usift}
\end{figure}

We use extended dynamical mean-field theory (EDMFT) to solve the model
at half-filling on the Bethe lattice with bandwidth $W=4$~eV, inverse temperature $\beta=40$~1/eV,
and a single bosonic mode $\omega_0=5$~eV and $\lambda=3.5$~eV. 
The spectral function and Selfenergy for different values
of the static interaction $U(\omega)$ is shown in Fig.~\ref{fig:spec_Usift}.
We observe the Hubbard satellite to completely vanish when reducing the interaction
by a constant shift,
where the quasiparticle peak recovers a renormalized semicircular form
that corresponds to the original non-interacting dispersion on the Bethe lattice,
renormalized only by the transfer of spectral weight into plasmon satellites.
This behavior if even more evident in the imaginary part of the Selfenergy
in Fig.~\ref{fig:spec_Usift} b). The low-energy peak responsible
for the Hubbard satellite is completely suppressed for vanishing $U(0)$, while the 
effect on the plasmonic peaks is small.
This confirms that the reduction of the interaction by a constant shift is
only effecting the plasmon satellites to a minor degree, as they originate
from the frequency dependence in $U(\omega)$.
On the other hand, the Hubbard satellites completely vanish at $U(\omega)=0$,
allowing for a systematic classification of the nature of the satellites.

\bibliographystyle{apsrev4-2}
\bibliography{materials,refs}